\documentclass[11pt,a4paper]{article}
\usepackage{jheppub}

\usepackage[T1]{fontenc}
\usepackage{amsmath,amsfonts,amsbsy,amssymb,enumerate,array}
\usepackage{lmodern}
\usepackage{verbatim,setspace}

\newcommand{\Scal}[1]{\Bigl ({#1} \Bigr )}
\newcommand{\scal}[1]{\bigl ({#1} \bigr )}
\def\ie{{\it i.e.}\ }
\def\eg{{\it e.g.}\ }

\DeclareMathAlphabet{\mathpzc}{OT1}{pzc}{m}{it}
\usepackage{dsfont}
\usepackage{amssymb}
\usepackage{amsmath}

\newcommand{\ord}[1]{{\scriptscriptstyle (#1)}}

\def\eq#1{(\ref{#1})}

\def\cN{{\mathcal{N}}}

\newcommand{\Pp}{\Phi_{\scriptscriptstyle +}}
\newcommand{\Pm}{\Phi_{\scriptscriptstyle -}}
\newcommand{\Ppm}{\Phi_{\scriptscriptstyle \pm}}
\newcommand{\Pmp}{\Phi_{\scriptscriptstyle \mp}}

\def\bea{\begin{eqnarray}}
\def\eea{\end{eqnarray}}
\def\be{\begin{equation}}
\def\ee{\end{equation}}

\newcommand{\CR}{\nonumber \\*}
\def\nn{\nonumber}
\def\cH{{\mathcal H}}

\newcommand{\pA}{{\text{\tiny A}}}

\def\cE{{\mathcal{E}}}

\def\ax{\alpha}

\usepackage{color}

\definecolor{cardinal}{rgb}{0.6,0,0}
\definecolor{darkgreen}{rgb}{0,0.4,0}
\definecolor{darkblue}{rgb}{0, 0, 0.7}

\makeatletter
\newcommand{\thickhline}{%
    \noalign {\ifnum 0=`}\fi \hrule height 1.3pt
    \futurelet \reserved@a \@xhline
}

\usepackage{datetime}

\title{Non-BPS multi-bubble microstate geometries}

\preprint{IPhT-T15/188,~\,CPHT-RR042.1015}

\author[a]{Iosif Bena,}  \author[b]{Guillaume Bossard,}  \author[c]{Stefanos Katmadas}  \author[a]{and  David Turton}
\affiliation[a]{
Institut de Physique Th\'eorique, Universit\'e Paris Saclay, CEA, CNRS, \\ 
\hspace{.14cm} 91191 Gif sur Yvette, France}
\affiliation[b]{Centre de Physique Th\'eorique, Ecole Polytechnique, CNRS, Universit\'e Paris-Saclay, \\
91128 Palaiseau Cedex, France}
\affiliation[c]{Dipartimento di Fisica, Universit\'a di Milano--Bicocca and INFN, Sezione di Milano--Bicocca,\\
Milano, Italy}

\emailAdd{iosif.bena[at]cea.fr}
\emailAdd{guillaume.bossard[at]cpht.polytechnique.fr} 
\emailAdd{stefanos.katmadas[at]unimib.it}
\emailAdd{david.turton[at]cea.fr}

\abstract{We construct the first smooth horizonless supergravity solutions that have two topologically-nontrivial three-cycles supported by flux, and that have the same mass and charges as a non-extremal D1-D5-P black hole.
Our configurations are solutions to six-dimensional ungauged supergravity coupled to a tensor multiplet,
and uplift to solutions of Type IIB supergravity.
The solutions represent multi-center generalizations of the non-BPS solutions of Jejjala, Madden, Ross, and Titchener, which have over-rotating angular momenta.
By adding an additional Gibbons--Hawking center, we succeed in lowering one of the two angular momenta below the cosmic censorship bound, and bringing the other very close to this bound.
Our results demonstrate that it is possible to construct multi-center horizonless solutions corresponding to non-extremal black holes, and offer the prospect of ultimately establishing that finite-temperature black holes have nontrivial structure at the horizon. 

}

\begin{document}

\maketitle


\section{Introduction}

The black hole information paradox~\cite{Hawking:1976ra} represents a	long-standing challenge for any theory of quantum gravity.
Over the past few years, following its sharpening using quantum information theory~\cite{Mathur:2009hf}, it has become increasingly clear that in order to solve this paradox there must be new physics at the black hole horizon.
There are many arguments that lead to the same conclusion, some focused on the experience of infalling observers~\cite{Almheiri:2012rt,Mathur:2012jk,Almheiri:2013hfa,Mathur:2013gua} (see also~\cite{Braunstein:2009my}), 
some based on the AdS-CFT correspondence \cite{Skenderis:2008qn,Kabat:2014kfa}, and some based on quantizing fields at the horizon \cite{Englert:2010cg}.

A common approach is to replace the vacuum at the black hole horizon by nontrivial structure that allows information to escape, thus preserving unitarity~\cite{Lunin:2001jy,Bena:2006kb,Brustein:2012sa,Dodelson:2015toa,Hawking:2015qqa}.\footnote{There are also approaches that involve postulating nonlocal physics outside black hole horizons~\cite{Giddings:2012gc,Papadodimas:2012aq,Maldacena:2013xja}.}
However, attempts to construct structure at the horizon face three challenges. First, horizons are null surfaces, and thus naive attempts to put structure there fail: massive objects fall through the horizon, while massless fields dilute in a horizon-crossing time. Second, black holes have entropy, so any structure replacing the horizon must have entropy large enough to reproduce the Bekenstein--Hawking entropy of the black hole. Third, the size of a black hole horizon increases as one increases Newton's constant, $G_N$, so that any kind of structure that might replace it should also grow with $G_N$ in exactly the same way.

The most successful approach to constructing such structure, passing all of the above tests, is the fuzzball programme in string theory~\cite{Mathur:2005zp,Bena:2007kg,Balasubramanian:2008da,Chowdhury:2010ct,Mathur:2012zp,Bena:2013dka}. In this programme one often considers semi-classical microstates, which are well-described within supergravity. The resulting supergravity solutions are known as ``microstate geometries'' or black hole solitons. These microstate geometries have no horizon or singularities, but have nontrivial topology supported by fluxes, such that the solutions have the mass and charges of a black hole. For large supersymmetric black holes a very large number of such microstate geometries have been constructed (see for example \cite{Bena:2005va,Berglund:2005vb,Bena:2006kb,Bena:2010gg,Lunin:2012gp,Giusto:2012yz,Bena:2015bea})
and their entropy has been argued to reproduce the growth with charges of the Bekenstein--Hawking entropy of the black hole \cite{Bena:2014qxa}. Similarly, one can also construct microstate geometries for extremal non-BPS black holes by starting from almost-BPS multi-center solutions \cite{Goldstein:2008fq} and performing certain duality transformations \cite{Dall'Agata:2010dy}. Thus, for extremal black holes, this programme has had considerable success.

However, non-extremal black holes present a much greater challenge. To date there exists only a handful of exact microstate solutions that have the mass and charges of non-extremal black holes. The earliest-known examples are the solutions of Jejjala, Madden, Ross and Titchener (JMaRT) \cite{Jejjala:2005yu} and their generalizations~\cite{Giusto:2007tt,AlAlawi:2009qe,Banerjee:2014hza}. 
The JMaRT solutions have more angular momentum than a physical black hole with the same mass and charges, and hence correspond to CFT states that are far away from the sector which dominates the black hole ensemble. In addition, these solutions have a single topologically-nontrivial cycle, and the methods originally used to find these solutions do not appear useful for constructing solutions with more complicated topologies. 

Other smooth non-extremal geometries have been found by generalizing the known systems describing extremal solutions, for example the running-Bolt solution \cite{Bena:2009qv} and its multi-center generalization \cite{Bossard:2014yta}. Unfortunately, it turns out that these solutions violate the BPS bound and do not admit a spin structure~\cite{Gibbons:2013tqa,Bossard:2014yta}, and hence they are not good candidates for describing black hole microstates. 
There is also a proposal for constructing microstates of neutral black holes~\cite{Mathur:2013nja}, which can be very long-lived, but do not appear to be described by stationary supergravity solutions.

Besides the above exact solutions, there exists a proposal to build large classes of microstate geometries for near-extremal black holes by placing negatively-charged probe supertubes in supersymmetric solutions \cite{Bena:2011fc,Bena:2012zi}. 
The action of these supertubes has metastable minima, but it has recently been shown that these solutions are classically unstable to decay into supersymmetric microstates~\cite{Bena:2015lkx}.


Given this state of affairs, it appears that the most promising direction towards building smooth microstate geometries with the asymptotic mass and charges of non-extremal black holes is to construct multi-centered generalizations of the JMaRT solution and similar solutions. 
The first step in this direction was the discovery, by two of the present authors, of a partially-solvable system of differential equations that describes solutions with non-extremal asymptotic structure, and includes the JMaRT solutions~\cite{Bossard:2014ola}. 

This system, described in detail in Section \ref{sec:sys-ans}, is built upon an auxiliary four-dimensional Euclidean Einstein--Maxwell subsystem, similar to that of~\cite{Bena:2009fi}. 
Thus, it allows one to start from a known gravitational instanton with a set of desirable properties and to construct solutions systematically. 
The four-dimensional instanton underlying the JMaRT solution contains a two-dimensional surface, known as a bolt. 
It turns out that it is straightforward to construct solutions with more topological cycles by starting from other instantons with Gibbons--Hawking centers at a finite distance from the bolt. 
In principle this method can be used to construct solutions with an arbitrary number of Gibbons--Hawking centers.

Our present goal is to construct black hole microstate geometries, which are {asymptoti\-cally-flat} solutions that have no horizons or closed timelike curves (CTCs) and are smooth up to acceptable singularities.\footnote{We emphasize the importance of constructing structure that replaces the black hole horizon using smooth horizonless solutions, that can be described in a controllable way. Singular solutions can vastly over-count the black hole entropy, and should therefore be discarded unless one can argue that they arise as limits of smooth solutions, or that one understands the mechanism by which the singularity is resolved in string theory. For further discussion, see~\cite{Bena:2013dka}.}
The purpose of this paper is to give a proof of principle of the possibility of constructing multi-center generalizations of the JMaRT solutions, and more generally, of constructing multi-bubble non-extremal black hole microstate geometries. We do this by considering its simplest extension, obtained by adding to the bolt a single Gibbons--Hawking center.

Our solution is the first smooth horizonless non-extremal black hole microstate geometry that has more than one topologically-nontrivial three-cycle. The solution has two three-cycles: the first is the three-dimensional bolt already present in the JMaRT solution, and the second extends between the bolt and the additional Gibbons--Hawking center, and is supported by nontrivial flux. 

In the JMaRT solutions, both angular momenta are over-rotating with respect to the black hole regime of parameters. In our solutions one angular momentum is within this bound, while the other exceeds it by a rather small amount. Thus our construction represents a significant improvement in this respect. We will discuss this in detail in due course.

The structure of this paper is as follows. In Section \ref{sec:sys-ans} we give a self-contained exposition of the system of \cite{Bossard:2014ola} in its six-dimensional incarnation, describing solutions of $\cN=(1,0)$ supergravity in six dimensions coupled to a single tensor multiplet, or of Type IIB supergravity compactified on $T^4$ or $K3$. We further present a class of solutions to this system, which in principle allows for an arbitrary number of Gibbons--Hawking  centers to be added to the JMaRT bolt. We then proceed in Section \ref{sec:reg-analysis} to perform a detailed analysis of the asymptotic structure, smoothness and absence of CTCs for a solution with a single additional Gibbons--Hawking center. These requirements lead to a number of algebraic constraints on the parameters of the solution, most of which can be solved explicitly, with three polynomial constraints remaining as nontrivial conditions to be satisfied. In Section \ref{sec:top-flux}, we discuss the topology of the smooth solution and the fluxes supporting it, commenting on the topology of solutions with more Gibbons--Hawking centers. In Section \ref{sec:num-example} we solve the three remaining polynomial constraints, and present an explicit set of parameters that gives a smooth microstate geometry. Section \ref{sec:disc} contains concluding remarks, and the two appendices describe the relation of our six-dimensional ansatz to five- and four-dimensional supergravity, and give the explicit expressions of the vector fields appearing in our solution.

\section{The ansatz for six-dimensional supergravity}
\label{sec:sys-ans}

We work in six-dimensional $\cN=(1,0)$ supergravity, coupled to a single tensor multiplet.
The field content of this theory is the metric, a two-form potential $B$, and a scalar $\phi$. The theory is a consistent truncation of Type IIB supergravity compactified on  $T^4$, and also of the $\cN=(2,0)$ effective six-dimensional supergravity describing Type IIB string theory compactified on $K3$. The two-form potential in six dimensions descends from the IIB Ramond-Ramond two-form, while the scalar field $e^{2\phi}$ can be viewed both as the dilaton and the warp factor of the internal $T^4 / K3$, since the two are equal in this truncation.

From a string theory point of view, our system describes a D1-D5-P bound state where the D1-branes wrap a circle with coordinate $y$, and D5-branes wrapping the $y$ circle and the internal $T^4 / K3$, and where the momentum charge P is along $y$. We consider the internal four-dimensional space to be microscopic, while the $y$ circle $S^1_y$ is macroscopic, and so our six-dimensional asymptotics are $\mathds{R}^{4,1} \times S^1_y$.
The resulting effective string in six dimensions carries both electric and magnetic charge with respect to $B$~\cite{Strominger:1996sh}.

To construct non-supersymmetric solutions to this theory, we use the partially-solvable system of differential equations discovered in \cite{Bossard:2014ola}, whose solutions automatically solve the equations of motion of supergravity. This system was found by considering the three-dimensional non-linear sigma model over a para-quaternionic symmetric space that one obtains after dimensional reduction of $\cN=(1,0)$ supergravity in six dimensions along one time-like and two space-like isometries. The relevant equations are given in terms of the Ernst potentials underlying the solutions to an auxiliary Euclidean Maxwell--Einstein subsystem (similar to other related systems \cite{Bena:2009fi, Bossard:2014yta}), which we now discuss.

All solutions to the four-dimensional Euclidean Maxwell--Einstein equations with one $U(1)$ isometry  can be described in terms of an $SL(3)/GL(2)$ non-linear sigma model coupled to Euclidean gravity in three dimensions, upon reduction along the isometry. The relevant degrees of freedom are the four Ernst potentials $\cE_\pm$ and $\Ppm$, which satisfy the equations 
\begin{eqnarray} \label{eq:EK-eoms}
\scal{ \cE_+ + \cE_- + \Pp  \Pm  } \Delta \cE_\pm   &=& 2 ( \nabla \cE_\pm +  \Pmp \nabla \Ppm ) \nabla \cE_\pm \,, \CR
 \scal{ \cE_+ + \cE_- + \Pp  \Pm  } \Delta \Ppm   &=& 2 ( \nabla \cE_\pm +  \Pmp \nabla \Ppm ) \nabla \Ppm \,.
\end{eqnarray}
The potentials determine the three-dimensional Riemannian metric $\gamma_{ij}$ via 
\begin{equation}\label{eq:R-base}
R(\gamma)_{ij} =  \frac{ ( \partial_{(i} \cE_+ + \Pm  \partial_{(i} \Pp  ) (  \partial_{j)} \cE_- + \Pp   \partial_{j)}  
\Pm  )}{( \cE_+  + \cE_- + \Pp  \Pm  )^2} - \frac{\partial_{(i} \Pp    \partial_{j)} \Pm  }{ \cE_+  + \cE_- + \Pp  \Pm  }    \,. 
\end{equation}
The four-dimensional metric is then determined by the potential $V$ and the vector $\sigma$, which are given by
\begin{equation} \label{DefVw}
V^{-1} =\, \cE_+ + \cE_- + \Pp  \Pm \ , \qquad
\star d\sigma = V^2\, (d \cE_+ - d \cE_- + \Pm  d \Pp  - \Pp  d \Pm )\  . 
\end{equation}
This four-dimensional metric does not appear explicitly in our Minkowski-signature six-dimen\-sional metric, however it will be convenient to use $V$ and $\sigma$ in the following.

\subsection{Six-dimensional metric}

In the partially-solvable system of \cite{Bossard:2014ola}, the Einstein-frame metric takes the form
\begin{equation}\label{eq:6D-metr}
ds^2 = \frac{H_3}{\sqrt{H_1 H_2}} ( dy + A^3)^2 - \frac{W}{H_3 \sqrt{H_1 H_2}} ( dt + k )^2 
       + \sqrt{H_1 H_2 } \Scal{ \frac{1}{W} ( d\psi + w^0 )^2 + \gamma_{ij} dx^i dx^j} \ , 
\end{equation}
where $\gamma_{ij}$ is the three-dimensional base of a solution to the Euclidean Maxwell--Einstein equations, as described above. Note that we write the metric in a form natural for a Kaluza--Klein reduction to five dimensions, where the relevant Kaluza--Klein vector field is $A^3$. The notation $A^3$ is motivated by the fact that it is one of the three gauge fields appearing symmetrically in the resulting five-dimensional theory. The vectors $A^3$ and $k$ decompose as
\begin{equation}\label{eq:6d-KK}
A^3 ~=~ A^3_t\, (dt + \omega) + \ax^3\,(d\psi + w^0) +  w^3 \,, \qquad\quad k ~=~ \omega + \frac{\mu}{W}\,(d\psi + w^0) \,.
\end{equation}
The expressions for the scalar $\mu$ and the vectors $\omega$, $w^0$, $w^3$ are given below, while the expressions of $A^3_t$ and $\ax^3$ are displayed in the next subsection in Eqs.\;\eqref{eq:5dzeta} and \eqref{eq:5dax} to emphasize the triality symmetry of the system. 

The ansatz is written in terms of three layers of functions. Firstly we have the four Ernst potentials underlying a solution to the Euclidean Maxwell--Einstein equations. Secondly we have four functions $L^a,\, K_a$, for $a=1,2$, that solve certain linear equations in the Maxwell--Einstein background. Thirdly we have two functions, $L^3,\, K_3$ that solve linear equations in the same background, with sources quadratic in $L^a,\, K_a$. The set of functions $W$, $\mu$, $H_I$ (for $I={1,2,3}$) appearing in the metric and gauge fields are given in terms of combinations of these 10 functions.

To write the ansatz, we split the index $I=(a,3)$, with $a=1,2$, and we introduce the $SO(1,1)$ invariant metric\footnote{This metric identifies the theory as the first in an infinite class of theories including $n$ minimally coupled tensor multiplets, for which a corresponding ansatz can be built using the expressions given in this section, upon extending $\eta_{ab}$ to an $SO(1,n)$ invariant metric.}
\begin{equation}\label{eq:STU-eta}
\eta_{ab}  = \left(\begin{array}{cc} 0\ &\ 1\\1\ &\ 0\end{array}\right) \,
\end{equation}
and its inverse $\eta^{ab}$. The functions $W$, $\mu$, $H_I$ are then given by
\begin{align}\label{eq:scal-facts}
 W = &\, \frac1{16}\,(L^3)^2 -\frac1{4}\,V\,K_1 K_2 K_3\, \Pm \,,
 \nonumber\\
 H_a = &\, \frac14\, \eta_{ab}L^b ( L^3 - V\, \Pm  K_c L^c) + \frac14\,( V \, \Pm  L^1 L^2 -K_3) K_a \,,
 \nonumber\\
 H_3 = &\, \frac14\,V \,(\cE_- + \Pp  \Pm ) \left( (1 - V\,\cE_+)\,K_1 K_2 - \cE_+ L^3 \right) + \frac14\,V \,\cE_+^2 \Pm  K_3 \,,
\CR
\mu = &\, -W\,\Pp  -\frac1{16}\,\left( 2\,(1 - V\,\cE_+)\, K_1 K_2 - \cE_+ L^3 \right)\, (K_3 + V\,\Pm  L^1 L^2 ) 
\CR
    &\, -\frac1{16}\,V\, \left(2\, \cE_+ \Pm  K_3 - (\cE_- + \Pp  \Pm )\,L^3 \right) \, K_a L^a \,.
\end{align}
Similarly, the vector fields $w^0$ and $w^3$ which appear in the metric are determined from the first-order equations: 
\begin{align} \label{wIEq-1} 
\star d w^0 =&\, \frac14\,d L^3 - \frac12\,V\,\Pm  K_a d L^a - \frac12\,V\,K_3\,\Pm  d \cE_+ 
+ \frac12\,V\,(L^3+V\,K_1 K_2)\,\Pm  d \Pp   
\CR  
&\,   + \frac14\,K_1 K_2\, (  d V + \star d \sigma )    \, , 
\CR
\star d w^3 =&\, \frac12\,V\,\left( V^{-1} d K_3 - d (\Pm  L^1 L^2 ) 
  +(K_a + \Pm  L_a)\,d L^a - L^a\, d (K_a + \Pm  L_a) +2\,K_3\,d \cE_+ \right)
\CR
&\, - V\,(L^3 + V\, K_1 K_2)\, d \Pp  + 2\,V\,L^1 L^2 \, d \Pm 
  + \frac12\, ( L^a  K_a + \Pm  L^1 L^2) \star d \sigma \, ,
 \end{align}
while the vector field, $\omega$, corresponding to the time fibration, is determined by
\begin{align}\label{eq:5d-k}
4\,\star d \omega = 
 &\, d\scal{ \Pp  L^3 -  \cE_+ K_3}  + V\, \cE_-( K_a d L^a - L^a d K_a ) -V\,\Pp  \Pm  d( K_a L^a)
\CR 
&\, +2\, V\,\cE_- \left( K_3\, d\cE_+ - ( L^3 + V\, K_1 K_2 ) d \Pp \right) + V\, K_a L^a \,d \cE_+ 
\CR
&\, + V\,\cE_+ \Pm   d( L^1 L^2) +V\,(\Pm  d\cE_+ - \cE_+ d \Pm ) L^1 L^2 
\CR
&\, -2\,V^2 \Pp  ( d\cE_- + \Pp  d \Pm ) K_1 K_2 - \cE_+ (K_a L^a + \Pm  L^1 L^2 ) \star d \sigma\, .
\end{align}

\subsection{The matter fields}

We next describe the ansatz for the matter content of the theory. Firstly, the scalar field, which can be identified with the dilaton of the D1-D5 system, is given by 
\begin{equation}\label{eq:6D-dil}
 e^{2\phi} = \frac{H_1}{H_2} \;. 
\end{equation}
The equation of motion for the two-form potential $B$, 
\begin{equation} 
d\left( e^{2\phi} \star_6 H \right) = 0  \,,
\end{equation}
expressed in terms of the three-form field strength, $H=dB$, can be recast by introducing the dual three-form field strength, $\tilde H = d \tilde B$, as
\begin{equation} \label{eq:6D-eom}
e^{\phi} \star_6 H +e^{-\phi} \tilde{H} = 0  \,. 
\end{equation}
The dual three-form $\tilde H = d \tilde{B}$ can be thought of as a magnetic dual to $H$, similar to the dual vector field strengths appearing in four-dimensional theories.
The two-form $B$ can be identified with the Ramond-Ramond two-form potential of the D1-D5 system in Type IIB supergravity on $T^4/K3$, whereas $\tilde{B}$  descends from the Ramond-Ramond 6-form wrapping $T^4/K3$.

The two-form potentials $B$ and $\tilde B$ can be expressed in terms of three-dimensional quantities. We first introduce the scalars $A_t^a$, $\beta_a$ and $\ax^a$, with the latter identified as two of the three axions in the reduction to four dimensions. We then introduce the three-dimensional one-forms $v_a$, $w^a$ and $b_a$, which will be defined shortly. Finally, we define the two-forms in three dimensions, $\Omega_a$, through
\begin{equation} \label{eq:Om-def}
d \Omega_a = v_a \wedge d w^0 - \eta_{ab} w^b \wedge d w^3 + b_a \wedge d \omega \,.
\end{equation}
In terms of these quantities, we have
\bea
B \,&=&\, A_t^1\, (dy + w^3 ) \wedge(dt+\omega) + \ax^1\, (dy + w^3)  \wedge (d\psi + w^0)
-\beta_2\,(dt+\omega)  \wedge (d\psi + w^0) \quad \CR
&&{} - w^1 \wedge (dy + w^3 )+ b_2 \wedge(dt+\omega) + v_2 \wedge (d\psi + w^0 ) + \Omega_2 \,,
 \nn
\eea
\vspace{-12mm}
\bea
\tilde B \,&=&\, A_t^2\, (dy + w^3 ) \wedge(dt+\omega) + \ax^2\, (dy + w^3)  \wedge (d\psi + w^0)
-\beta_1\,(dt+\omega)  \wedge (d\psi + w^0) \quad \CR
&&{} - w^2 \wedge (dy + w^3 )+ b_1 \wedge(dt+\omega) + v_1 \wedge (d\psi + w^0 ) + \Omega_1 \,.
\label{eq:2-form-exp}
\eea
%
%
Note that the $\Omega_a$ ensure that in $H$ and $\tilde{H}$, the vectors $w^a$, $b_a$ and $v_a$ only appear through the gauge-invariant quantities $dw^a$, $db_a$ and $dv_a$.
The $\Omega_a$ vanish for axisymmetric solutions, since all vector fields have components along the angular coordinate around the axis, implying that their wedge products appearing in \eqref{eq:Om-def} vanish identically. We only construct axisymmetric solutions in the current work, so we now set $\Omega_a$ to zero.

The one-forms, $w^a$, $v_a$, $b_a$ in \eqref{eq:2-form-exp} are determined in terms of the functions appearing in the ansatz  by solving the first-order equations
\begin{align}
\label{wIEq-2} 
  \star d w^a ~=~ &\, \frac12\,d \Big( \eta^{ab}K_b - \cE_+ V\, (\eta^{ab}K_b + \Pm  L^a) \Big)  + \cE_+ V\, L^a d \Pm  
\CR
 &\,- \cE_+ V^2\,(\eta^{ab}K_b + \Pm  L^a) (d \cE_- + \Pp  d \Pm  ) \   ,
\\
\label{baEq-2} 
\star db_a ~=~&\, 
V\, (\eta_{ab} \,\Phi_{-} d L^b + d K_a) - \eta_{ab}\,V\, L^b d\Phi_{-}  - (\eta_{ab} \,\Phi_{-}\, L^b + K_a) \star d \sigma\,,
\\
\label{eq:alm-NE-mag}
 \star d v_a ~=~ &\, -2\,\eta_{ab}L^b \, V\,d \cE_- - V\,d\left( \Pp  K_a - (\cE_+ + \cE_-) \eta_{ab}L^b \right) 
 \CR
  &\,  + \left( \Pp  K_a - (\cE_+ + \cE_-) \eta_{ab}L^b \right) \star d \sigma 
\,.
\end{align}
Their explicit form can be obtained straightforwardly for any given solution to the system. The scalars $\beta_a$ are given by
\begin{align}
\beta_1 = &\, -\frac{1}{2\,H_2} \left( K_3 + V \,(K_1 L^1 - K_2 L^2) - V \,\Phi_{-} L^1 L^2\right) ,
\CR
\beta_2 = &\, -\frac{1}{2\,H_1} \left( K_3 + V \,(K_2 L^2 - K_1 L^1) - V \,\Phi_{-} L^1 L^2\right)
.
\end{align}
Finally, the electric components $A^I_t$, of the five-dimensional vectors $A^I$ are given by
\begin{align}\label{eq:5dzeta}
 A^1_t = &\, \frac1{4\,H_1}\,\left( L^3 - 2\, V\, \Pm \,K_2 L^2 \right)\,,
 \CR
 A^2_t = &\, \frac1{4\,H_2}\,\left( L^3 - 2\, V\, \Pm \,K_1 L^1 \right)\,,
 \CR
 A^3_t= &\, \frac{1}{4\,H_3}\,\left((2\,V\,\cE_+ - 1) L^3 + 2\,(V\,\cE_+ - 1)\,V\,K_1 K_2 +2\,V\,K_3 \Pm \cE_+   \right)\,, 
\end{align}
while the three axions are 
\begin{align}\label{eq:5dax}
\ax^1 = &\, \frac{1}{4\,H_1}\,\left( V\,\cE_- (K_1 L^1 - K_2 L^2) - \Pp  L^3 + \cE_+ \,K_3 
         - V\, \left(\cE_+ \Pm  L^1 L^2 - \Pp \Pm  K_a L^a \right) \right)\,,
\CR
\ax^2 = &\, \frac{1}{4\,H_2}\,\left( V\,\cE_- (K_2 L^2 - K_1 L^1) - \Pp  L^3 + \cE_+ \,K_3 
         - V\, \left(\cE_+ \Pm  L^1 L^2 - \Pp \Pm  K_a L^a \right) \right)\,.
\CR
\ax^3 = &\, \frac{1}{4\,H_3}\,\left( \rule[-.3\baselineskip]{0pt}{\baselineskip}
   (1 - 2\, V\, \cE_+) \Pp L^3 + (1 - 2 V \Pm \Pp) \cE_+ K_3 + \cE_+ \Pm V L^1 L^2 \right.
\CR   &\,\quad\quad\quad \left.
         - (1 - V \cE_+) (K_a L^a - 2 V \Pp K_1 K_2) \rule[-.3\baselineskip]{0pt}{\baselineskip} \right)\,.
\end{align}
Note that in the above we have given the components $A^3_t$ and $\ax^3$ of the gauge field $A^3$ in \eqref{eq:6d-KK}, using a naming convention that highlights the triality that arises when the reduction to five- and four-dimensional supergravity is performed. In Appendix \ref{app:reduction} we give some details on the dimensional reduction of this solution to lower dimensions.

This ansatz is rather complicated, but is solvable by construction. The equations of motion satisfied by the Ernst potentials $\cE_\pm,\, \Ppm$ and the Euclidean three-dimensional base metric are displayed in \eqref{eq:EK-eoms}--\eqref{eq:R-base}. The six functions $L^I$ and $K_I$ solve a hierarchy of linear equations defined by the Bianchi identities for the vectors $dw^0$, $dw^I$, $d\omega$, $dv_a$ and $db_a$ in \eqref{wIEq-1}, \eqref{eq:5d-k}, \eqref{wIEq-2}, \eqref{baEq-2} and \eqref{eq:alm-NE-mag}. Once these functions are obtained, the solution is completely determined.

\subsection{Multi-center solutions}

We now turn to particular solutions to the system of the previous subsection. We first choose a Euclidean Einstein--Maxwell base, which
defines the three-dimensional base metric and the Ernst potentials appearing throughout the system of equations. 
As mentioned above, we will allow for extra poles in the Ernst potentials, describing Gibbons--Hawking--like centers, however we take the three-dimensional metric to be that of Euclidean Kerr--Newman throughout the paper.
This ensures the absence of conical singularities in the three-dimensional base metric. Such singularities are related to attractive net forces between the centers, which vanish in our ansatz.

It will be convenient to use spherical coordinates $(r, \theta, \varphi)$ in which the base metric takes the form
\begin{eqnarray}  \label{3Dbase} 
\gamma_{ij} dx^i dx^j &=& 
\scal{ r^2 - c^2+a^2 \sin^2\theta} \left( \frac{dr^2}{r^2 - c^2  } + d\theta^2 \right) 
+  \scal{ r^2 - c^2 }\, \sin^2\theta\, d\varphi^2 \,.
\end{eqnarray}
We can also express the metric in Weyl coordinates, defined through\footnote{Note that we have interchanged the definitions of $r_+$ and $r_-$ with respect to those of \cite{Bossard:2014ola}; in our conventions $r_+$ vanishes at the North Pole.}
\begin{equation}
r_\pm = \sqrt{ \rho^2 + (z\mp c)^2 } \ , \qquad 2\, r = r_+ + r_- \  , \qquad 2\, c\, \cos\theta = r_- - r_+ \,, 
\end{equation}
in terms of which 
\begin{eqnarray}  \label{3Dbase-2} 
\gamma_{ij} dx^i dx^j &=&
\frac{ r^2 - c^2 + a^2 \sin^2{\theta}}{r_+ r_- }\,( dz^2 + d\rho^2) + \rho^2\, d\varphi^2 \,.
\end{eqnarray}

The fact that we impose a particular three-dimensional base, rather than solving for it through \eqref{eq:R-base}, implies additional restrictions on the Ernst potentials. We therefore consider the following ansatz for these potentials,
\begin{align}\label{eq:Kerr-mod}
 \cE_+ =&\, -1 + 
 \frac{2\,(r + a\, \cos{\theta}) }
              { r + a\, \cos{\theta} + m_+  + \mathcal{H}\, (r + a \cos{\theta} + \frac1{m_-}\,(c^2 - a^2) ) }\,,
\CR
 \cE_- =&\, 1 - \frac{ 2\, m_- }{ r - a \cos{\theta} + m_- }\,,
\CR
\Pp  =&\, \frac{1}{e_-}\,
 \frac{  m_+ m_-  - (c^2 - a^2)  +m_-\,\mathcal{H} \left( r + a \cos{\theta} + \frac1{m_-}\,(c^2 - a^2) \right) }
      { r + a \cos{\theta} + m_+  + \mathcal{H} \left( r + a \cos{\theta} + \frac1{m_-}\,(c^2 - a^2) \right) }\,,
\CR
 \Pm  =&\, \frac{ 2\,e_- }{ r - a\, \cos{\theta} + m_- }\,.
\end{align}
It was shown in \cite{Bossard:2014ola} that these potentials solve Eqs.\;\eqref{eq:EK-eoms} and \eqref{eq:R-base} provided that $\mathcal{H}$ is a solution to the following equation on the base:
\begin{equation} \label{eq:Lapl-H}
\Delta {\mathcal{H}} =  \frac{  2\,( c^2 -a^2)\, (r - a \cos{\theta} + m_{-}) }{(r^2-c^2 + a^2 \sin^2{\theta} ) \scal{m_{-} (r + a \cos{\theta} ) + c^2-a^2 }} \nabla (r + a \cos{\theta}) \cdot \;\!\! \nabla {\mathcal{H}} 
\end{equation}
This is a linear equation which is straightforward to solve, and its solutions can be superposed:
\begin{equation}\label{eq:H-gen}
\mathcal{H} = h +\sum_\pA \cH_\pA  \ ,
\end{equation}
where $h$ is a constant. The functions $\cH_\pA$ have poles at additional centers that we denote by $x_\pA$. We will only consider axisymmetric solutions, for which the additional poles are all on the rotation axis of the original Kerr--Newman solution, so that in Weyl coordinates we have $(z_\pA,\rho_\pA)=(R_\pA,0)$. Then $\cH_\pA$ takes the form
\begin{equation}\label{eq:H-one}
\hspace{-2mm} 
\cH_\pA  \,=\,  \frac{8\, n_\pA }{(R_\pA-a)\,( r + a \cos{\theta} + \frac{c^2-a^2}{m_-} )} 
\frac{ (R_\pA-a) r + (a\,R_\pA- c^2)\, \cos \theta}{\sqrt{ (R_\pA-r\cos\theta)^2 + (r^2 - c^2) \sin^2\theta}}\,,
\end{equation}
where $n_\pA$ are constants parametrizing the residue of the function $V$ at the poles. 

In terms of the Einstein-Maxwell theory, the Ernst potentials \eqref{eq:Kerr-mod} describe the Wick rotation of a Kerr-Newman black hole, when $\mathcal{H}=0$. In these coordinates, the horizon is at $r=c$ and remains a special locus in the full supergravity solution, as we will discuss in Section \ref{sec:nut-bolt-reg}. The extra poles in $\mathcal{H}$ can be viewed as describing Gibbons--Hawking--type centers, as can be verified by expanding the solution in their vicinity. 

Given this base, one can solve the Bianchi identities for the vector fields $dw^0$, $dw^I$, $d\omega$, $dv_a$ and $db_a$, given in \eqref{wIEq-1}, \eqref{eq:5d-k}, and \eqref{wIEq-2}--\eqref{eq:alm-NE-mag}, to obtain the functions $K_I$ and $L^I$. It was shown in \cite{Bossard:2014ola}  that a particular solution to this system can be defined in terms of the Ernst potential themselves as 
\begin{align}\label{eq:LK-gen}
 K_a = &\, \eta_{ab}q^b + (\cE_+ + 1) \eta_{ab}l^b\,,
\CR
 L^a = &\, p^a  - \left(\cE_+ + 1 - V^{-1} \right) \,\frac{l^a}{\Pm}\,,
\CR
 K_3 = &\, \left( \cE_+ + 1 - V^{-1}  \right)^2 \frac{V}{\Pm}\,l^1 l^2 + p^3\,V
 - \left( \frac{l^3}{\Pm} - q^3 \right) \,\left( V\,( \cE_+ + 1) - 1  \right) \,,
\CR
 L^3 = &\, - l^a K_a \,\left( V\,( \cE_+ + 1) - 1  \right)  - (p^3  \Pm + q^1 q^2)\,V 
 +\left( l^3 - q^3\,\Pm \right)\,V\,( \cE_+ + 1) \,,
 \end{align}
where $l^I$, $p^I$ and $q^I$ for $I=\{1,2,3\}$ are integration constants. The $p^I$ and $q^I$ are related to the asymptotic charges, while the $l^a$ parametrize the asymptotic values of the dilaton \eqref{eq:6D-dil} and the $g_{yy}$ component of the metric.

The above equations specify the solution for all the supergravity fields. Note that we have been able to write the entire solution in terms of the functions $\cE_+$ and $V$ appearing in the Maxwell--Einstein instanton. 
This is an artifact of the solution \eq{eq:LK-gen} representing a restricted ansatz and not the most general solution to the system. Furthermore, it will turn out that this ansatz cannot be used to construct smooth solutions with more than one additional Gibbons--Hawking center. Nevertheless, we will see in the following that it includes smooth microstate geometries with two nontrivial three-cycles.

\section{Regular solutions to the system}
\label{sec:reg-analysis}

From now on we will focus on solutions in which the function $\mathcal{H}$ has a single pole.
In order to simplify the required manipulations, it is useful to make some gauge transformations and coordinate transformations on the solution obtained by directly substituting \eqref{eq:LK-gen} in the relevant expressions. 

Firstly and most importantly, we shift away various asymptotic constants from the components of the metric and two-forms $B$ and $\tilde B$, using diffeomorphisms and gauge transformations respectively. Specifically, shifting the asymptotic values of the scalars $\ax^a$, $\beta_a$ and $A_t^a$ to zero in \eqref{eq:2-form-exp} is equivalent to a gauge transformation on the two-forms, provided that the vector fields are redefined as
\begin{align}\label{eq:redef1}
 w^a \rightarrow &\, w^a + A_t^a \big|_{\scriptscriptstyle\infty}\omega + \ax^a \big|_{\scriptscriptstyle\infty} w^0\ , 
\CR
 v_a \rightarrow &\, v_a - \beta_a \big|_{\scriptscriptstyle\infty} \omega 
                         + \eta_{ab}\,\ax^b \big|_{\scriptscriptstyle\infty} w^3 \ , 
\CR
 b_a \rightarrow &\, b_a + \eta_{ab}\,A_t^b \big|_{\scriptscriptstyle\infty} w^3
                         + \beta_a \big|_{\scriptscriptstyle\infty} w^0\ , 
\end{align}
where we denoted the asymptotic values of the scalars by $\big|_{\scriptscriptstyle\infty}$. In addition, one may remove the asymptotic constants of $A_t^3$ and $\ax^3$ appearing in the Kaluza--Klein gauge field $A^3$ in \eqref{eq:6d-KK} by a diffeomorphism mixing the coordinate $y$ with $t$ and $\psi$ at infinity, provided one imposes the redefinition
\begin{align}\label{eq:redef2}
 v_a \rightarrow &\, v_a + \ax^3 \big|_{\scriptscriptstyle\infty}\eta_{ab}\, w^b \ , 
\CR
 b_a \rightarrow &\, b_a + A_t^3 \big|_{\scriptscriptstyle\infty} \eta_{ab}\,w^b  \ , 
\CR
 \beta_a \rightarrow &\, \beta_a +\ax^3\big|_{\scriptscriptstyle\infty} \eta_{ab}\, A_t^b\ . 
\end{align}
Finally, we shift away the constant values of $\omega$, $w^3$ and the $w^a$ at infinity by appropriate mixing of the coordinates $t$, $y$ with $\varphi$ and a further gauge transformation on the two-forms respectively, which do not induce any additional redefinitions. Henceforth, we assume that the transformations \eqref{eq:redef1} and \eqref{eq:redef2} have been applied on all fields. The relevant asymptotic constants appearing are not illuminating and play no role in the following, so we refrain from giving them explicitly.

For later convenience we reparametrize the constant $n_1$ appearing in $\mathcal{H}$ via \eqref{eq:H-one} by $n_1=\tfrac{N}{q^1 q^2}$. We also set the asymptotic constant $h=-1$, which is required for asymptotic flatness as we will discuss next.
Then the function $\mathcal{H}$ in \eqref{eq:H-gen} becomes
\begin{equation}\label{eq:H-fin}
\mathcal{H} = -1 + \frac{8\,N}{q^1 q^2 (R-a)\left( r + a \cos{\theta} + \frac{c^2-a^2}{m_-} \right)} 
\frac{ (R-a) r + (a\,R- c^2)\, \cos \theta}{\sqrt{ (R-r\cos\theta)^2 + (r^2 - c^2) \sin^2\theta}}  \ ,
\end{equation}
where $R$ denotes the distance along the positive $z$ axis from the origin to the Gibbons--Hawking center. With this normalization, $N$ will turn out to be quantized as an integer when we impose smoothness.

\subsection{General requirements for asymptotic flatness and regularity} \label{sec:reg-general}

As we have discussed, to obtain a D1-D5-P black hole microstate geometry, we require $\mathds{R}^{4,1}\times S^1$ asymptotics, smoothness, and no closed timelike curves (CTCs). 
As found in \cite{Bossard:2014ola}, for $\mathds{R}^{4,1}\times S^1$ asymptotics, the parameters in \eqref{eq:LK-gen} are given by 
\begin{gather}
{l^I}  = 0\,,
\qquad
h = - 1 \,,
\qquad
e_- = \frac{1 + x}{q^3}\,,
\CR
p^1 = - 1 - \frac{m_-}{2\, (1 + x)}\,q^1 q^3\,,
\qquad
p^2 = - 1 - \frac{m_-}{2\, (1 + x)}\,q^2 q^3\,,\qquad
{ p^3} = \frac{m_- }{2\, (1 + x)}\,q^1 q^2 q^3\,,
\CR
m_+  = \frac1{m_-}\,(c^2-a^2 ) - \frac{4}{q^1 q^2}\,(x-1) - \frac{8}{q^1 q^2}\,N\,, \label{eq:par-JMaRT}
\end{gather}
where $x$ is a constant parametrizing $e_-$ that we introduce for convenience. The parameter $m_-$ is also fixed:
\begin{equation}\label{eq:mm-val}
 m_- = \frac1{4}\,(c^2-a^2 ) \,q^1 q^2 - \frac{q^1 + q^2}{q^1 q^2 q^3}\,(x^2-1)  - \frac{(1 + x)^2}{(q^3)^2} \,.
\end{equation}
We will mostly avoid using this explicit expression for ease of notation. With these choices, the various functions behave asymptotically as
\begin{gather}
 W =\frac{1}{r^{2}} + {\cal O}(r^{ -3})\,, 
 \qquad 
 H_I = \frac{1}{r} + {\cal O}(r^{-2})\,, 
\qquad w^0 = - \cos\theta d\varphi +  {\cal O}(r^{ -1})\ , 
 \CR
 \frac{\mu}{W} = \frac{-J_\psi + J_\varphi \, \cos{\theta}}{8 \, r} + {\cal O}(r^{-2})\,,
 \qquad
 \omega = -\frac{J_\varphi\, \sin^2{\theta}}{8 \, r} d\varphi+ {\cal O}(r^{ -2})\,,
 \label{eq:asym-flat}
\end{gather}
where $J_\psi$ and $J_\varphi$ stand respectively for the angular momenta along the directions $\psi$ and $\varphi$, and are given in \eqref{eq:MJ-gen} below. 
The coordinates $y$, $\psi$, $\varphi$ are subject to the identifications
\bea \label{eq:identifications}
y ~\sim~ y+2\pi R_y \,,\qquad  \psi ~\sim~ \psi+4\pi \,, \qquad 
(\psi, \varphi )~\sim~(\psi, \varphi )+ (2\pi, 2\pi) \,,
\eea
where $R_y$ will be fixed in terms of other parameters in the solution in due course.

Note that we set all $l^I=0$, whereas strictly speaking only $l^3=0$ is required to ensure that the spacetime be asymptotically $\mathbb{R}^{1,4}\times S^1$. These additional conditions moreover imply the dilaton $e^{2\phi}$ and $g_{yy}$ to tend to $1$ at asymptotic infinity. 
There is no loss of generality in doing this, since we keep the radius of the $y$ circle explicitly as $R_y$, and since more general asymptotic values of $e^{2\phi}$ can be obtained straightforwardly by an appropriate rescaling.

Because the solution describes a microstate of a five-dimensional black hole, it is useful to compute its five-dimensional asymptotic charges. The five dimensional solution (obtained by reduction on the asymptotic circle) carries three total electric charges:
\begin{align}\label{eq:Q-jmart}
Q_I = &\,4 \frac{ x^2-1 }{q^{I+1} q^{I+2}} - (a^2 - c^2)\,q^{I+1} q^{I+2}\,,
\end{align}
where $Q_a$ for $a=1,2$ are defined in six dimensions as the asymptotic fluxes of the three-form $H$ and its dual $\tilde{H}$, as shown in Section \ref{sec:top-flux}, and $R_y^{\, 2} Q_3$ represents the asymptotic momentum along the $y$ direction.

The five-dimensional ADM mass and the angular momenta along the remaining two directions, $\psi$ and $\phi$, are given by\footnote{We have used the explicit expression for $R$ that will come later in \eqref{eq:Rsol} to simplify the second factor in $J_\varphi$.}
\begin{align}\label{eq:MJ-gen}
M_{\mbox{\tiny ADM}}= &\,  \sum_I E_I\,, \qquad 
\CR
 J_{\varphi} = &\, a \,\biggl( (a^2- c^2)\,q^1 q^2 q^3 + 4 (x^2 - 1) \sum_I \frac{1}{q^I} \biggr) + \frac{16 N(x+1)}{q^1 q^2 q^3} \frac{(R^2 -c^2)}{(R-a)^2} \,, 
\CR
 J_{\psi} = &\,2\, x \biggl( (a^2- c^2)\,\sum_I q^I + 4 \frac{x^2 -1}{q^1 q^2 q^3} \biggr) \,,
\end{align}
where the constants $E_I$ are given by\footnote{Note the redefinition with a factor of 4 with respect to \cite{Bossard:2014ola} in order for $E_I$ to coincide with the charges in the BPS limit.}
\begin{align}\label{eq:E-jmart}
E_I = &\, 4\, \frac{ x^2-1 }{q^{I+1} q^{I+2}} + ( a^2 - c^2 )\,q^{I+1} q^{I+2}\,,
\end{align}
and satisfy the conditions
\begin{equation} \label{ConE} 
 E_I^2 =  Q_I^2 +16 (x^2-1)\,(a^2 - c^2)\,.
\end{equation}

Our solution has coordinate singularities at the bolt and the additional center at the pole of $\cH$ in \eqref{eq:H-fin}. To ensure smoothness, we must therefore show that various functions have poles of the usual type compatible with regularity of the full metric. 
We first analyze the conditions for regularity away from such special points, postponing an explicit discussion of these for the next subsection. 

The determinant of the metric is 
\be 
g = H_1 H_2 (r^2 - c^2+a^2\sin^2\theta)^2 \sin^2\theta\,
\ee
and so away from special points, the functions $H_1$ and $H_2$ cannot go to zero or infinity. Given their $1/r$ behaviour at infinity, as in \eqref{eq:asym-flat}, it follows that $H_1$ and $H_2$ must be strictly positive and finite everywhere away from the special points.

To find the conditions for the absence of CTCs, we take the line element and complete the squares successively in $y$, $\psi$, $\varphi$. The metric in these periodic directions must not have negative eigenvalues, so the three diagonal terms must be non-negative.

After completing the squares in $y$ and $\psi$, the line element then takes the form
\bea
ds^2 &=& \frac{H_3}{\sqrt{H_1 H_2}} ( dy + A^3)^2 
+ \frac{H_1H_2H_3-\mu^2}{WH_3\sqrt{H_1H_2}}
\left[(d\psi+w^0)-\frac{W\mu}{H_1H_2H_3-\mu^2}(dt + \omega)\right]^2 \cr
&&{}
 + \sqrt{H_1 H_2 } \left[ - \frac{W}{H_1H_2H_3-\mu^2} ( dt + \omega )^2  + \gamma_{ij} dx^i dx^j\right].
\eea
Firstly, considering the $g_{yy}$ component of the metric, we see that $H_3$ must be positive.
Similarly, we see that from the prefactor of the $\psi$ fiber combination that we require 
\bea \label{eq:cond1}
\frac{H_1H_2H_3-\mu^2}{W} &\ge& 0 \,.
\eea
Finally, using the form of the 3D base metric \eq{3Dbase} we complete the square on $d\varphi$, obtaining the prefactor
\bea
\sqrt{H_1 H_2 } \left[ - \frac{W}{H_1H_2H_3-\mu^2} \omega_\varphi^2  + (r^2-c^2)\sin^2\theta \right],
\eea
which gives the condition 
\bea \label{eq:cond2}
(r^2-c^2)\sin^2\theta &\ge&
 \frac{W}{H_1H_2H_3-\mu^2} \omega_\varphi^2 \,.
\eea
Note that this implies that $\omega$ must vanish when the left-hand side is zero, or when $W/(H_1H_2H_3-\mu^2)$ has a pole. Using \eqref{eq:asym-flat}, one finds that $(H_1H_2H_3-\mu^2)/W \to 1/r$ as $r\to \infty$, so $\omega$ must vanish at $r \to \infty$. All together, $\omega$ must vanish at $r=c$, at $\sin\theta = 0$, and as $r \to \infty$. 

In view of the fact that the regularity conditions are given explicitly in terms of the values of the various vector fields, we present the explicit form of these fields throughout the solution in Appendix \ref{app:vecfields}. These were obtained by using the expressions \eqref{eq:LK-gen} in the relevant ansatze \eqref{wIEq-1}, \eqref{eq:5d-k}, \eqref{wIEq-2} for the vector fields and imposing the redefinitions \eqref{eq:redef1}--\eqref{eq:redef2} above. All values of vector fields appearing below should be understood to be obtained from the expressions in the Appendix, upon taking the appropriate limits.

\subsection{Regular nuts and bolts}
\label{sec:nut-bolt-reg}

Given the complexity of the explicit solution, we proceed in two steps. Firstly we investigate the solution analytically around the special points, and secondly we analyze explicit examples of the parameters, to show that it is possible to obtain everywhere smooth solutions. The explicit examples will be discussed in Section~\ref{sec:num-example}; we now begin the analytic investigation.

The special points are characterized by the loci where some of the three $U(1)$ isometries of the solution, corresponding to the Killing vectors $\partial_y$, $\partial_\psi$ and $\partial_\varphi$, degenerate. The local geometry around such loci can be made regular upon imposing appropriate conditions on the metric, so that they can be viewed as smooth origins of certain subspaces. In order to study these special regions, we consider a time-like slice of the full six-dimensional metric \eqref{eq:6D-metr} and a Killing vector, $K$, assumed to be a linear combination of the three $U(1)$ isometries above. 

Following \cite{Gibbons:1979xm}, the locus $K^2=0$ describes a set of fixed points of the isometry, which can be characterized by considering the action of the isometry on the tangent space at the given point. This action is generated by $\nabla_a K_b \equiv \nabla_{[a} K_{b]}$ which, assuming it is nontrivial, must have rank two or four in five Euclidean dimensions.

When $\nabla_a K_b$ has rank four, there will be a one-dimensional subspace that is invariant under the action of the $U(1)$ isometry $K$. The locus $K^2=0$ is an isolated point in the remaining four directions, so that the local geometry is that of (a smooth discrete quotient of) $\mathds{R}^4\times S^1$, near the origin of $\mathds{R}^4$. We call this a nut, as it is the straightforward uplift on a circle of a standard nut in four Euclidean dimensions \cite{Gibbons:1979xm}. Note, however, that there exist smooth discrete quotients of $\mathds{R}^4\times S^1$ where the orbifold singularity present in $\mathds{R}^4$ is resolved in the total space. This is precisely what happens at the special points of our solution.

Similarly, when $\nabla_a K_b$ has rank two, there will be a three-dimensional subspace that is invariant under the action of the $U(1)$ isometry $K$. Then $\nabla_a K_b$ only acts nontrivially on a two-dimensional subspace of the tangent space at the fixed point, so that the local geometry is that of a product of $\mathbb{R}^2$ times a three-dimensional invariant compact submanifold. We call this a bolt, after the corresponding rank-two fixed point appearing in four Euclidean dimensions \cite{Gibbons:1979xm}. In this paper, we will only deal with simply-connected three-cycles, so that the local geometry near the bolt is a smooth discrete quotient of $\mathbb{R}^2\times S^3$, but other possibilities, for example $S^2\times S ^1$, may also exist. 

Note that the notion of a bolt introduced above is based on whether this locus is a fixed point of a single $U(1)$ isometry, without any reference to possible additional isometries. It is common to distinguish the  $U(1)$ isometries that define the structure of a solution from accidental isometries that may occur in more restricted classes of solutions. For example, a general Gibbons--Hawking instanton with $N$ nuts only admits one $U(1)$ isometry. It is only when all centers lie on the same axis that the solution admits an extra $U(1)$ isometry; this then defines fixed loci between the centers which are bolts in the sense defined above. However, it is common terminology not to refer to these as bolts, since they are an artifact of the additional $U(1)$ isometry. This is a general feature of axisymmetric solutions that have a flat three-dimensional base metric, and which generally admit non-axisymmetric generalizations.  Those include for instance the supersymmetric limit of the JMaRT solution, for $a=c$~\cite{Giusto:2004id,Giusto:2004ip}, whose three-dimensional base metric is flat~\cite{Giusto:2004kj}. Therefore we will not refer to its $S^3$ bubble as a bolt. On the contrary, when the bolt locus is defined at a conical singularity of the three-dimensional base metric, the two singular behaviors compensate each other to define a regular four or five-dimensional Euclidean metric and the degenerating $U(1)$ isometry is absolutely essential in describing the local geometry. In practice, one only calls a bolt a degenerate locus that is in the second category.

The existence of such a bolt is a general feature of gravitational instantons originating from non-extremal black hole solutions by analytic continuation to Euclidean signature. The Killing horizon is by definition a codimension-two surface where the norm of the Killing vector vanishes, such that after analytic continuation it leads to the singular locus of an isometry -- a bolt -- provided the original black hole had a nontrivial surface gravity. 
The three-dimensional base metric we use is that of a Euclidean non-extremal Kerr--Newman black hole, described by the Ernst potentials \eqref{eq:Kerr-mod} for $\mathcal{H}=0$.
Thus the solutions described in this paper admit a Killing vector with a nontrivial bolt homeomorphic to (the discrete quotient of) a three-sphere at $r=c$. 
This is the case irrespectively of the number and positions of the extra poles of the function $\mathcal{H}$. The fact that the bolt is naturally associated to a Killing horizon with non-zero temperature in Euclidean gravity is a sign that smooth solitons admitting a bolt are associated to non-extremal black hole microstates.

For the metric \eqref{eq:6D-metr}, the relevant vector field $K_B$ collapsing on the bolt is of the type 
\begin{equation}\label{KillBolt} 
 K_B= R_y \partial_y  + (m+n) \partial_\psi - (m-n) \partial_\varphi \;, 
\end{equation}
where $R_y$ is the radius of the $y$ circle and $m$ and $n$ are integers. For constant $\mathcal{H}$ this bolt is the unique cycle of the solution and, as was shown in \cite{Bossard:2014ola}, the solution reduces to the JMaRT solution \cite{Jejjala:2005yu}. 

In addition, we find that two linear combinations of the three $U(1)$ isometries collapse at the two poles of the bolt and at the Gibbons--Hawking center located at the pole of $\mathcal{H}$. Thus all three special points of our solutions are nuts. We now proceed to discuss in turn the geometry near the nuts and the bolt.

\subsection{Geometry at the Gibbons--Hawking center}

The Gibbons--Hawking center is located at $r_1=0$ in the coordinates 
\be 
r_1 = \, \sqrt{ (R-r\cos\theta)^2 + (r^2 - c^2) \sin^2\theta} \ ,  \qquad\quad \cos \theta_1 = \, \frac{r \cos \theta - R}{r_1} \ . 
\ee
The limits of the functions $W$ and $H_I$ are
\begin{align}\label{eq:extr-beh}
 W =\frac{N^2}{r_1^{\; 2}} + {\cal O}(r_1^{\; -1})\,, 
 \qquad \quad 
 H_I = h_I^E\,\frac{N}{r_1} + {\cal O}(r_1^{\; 0})\,,
\end{align}
where we define the constants
\begin{equation}  
h_I^E \equiv \left(1 - \frac{2\,(x+1)}{R-a}\,\frac{1}{q^{I+1} q^{I+2} } \right) \ , 
\end{equation}
which must all be strictly positive.
Given the behavior in \eq{eq:extr-beh}, the absence of closed-time-like curves requires from \eqref{eq:cond1} and \eq{eq:cond2} that
\begin{equation}
\frac{\mu}{W}\bigg|_{r_1=0} ~=~ 0 \,,  \qquad\quad \omega\big|_{r_1=0} ~=~ 0  .
\end{equation}
Each of these two conditions independently determines the distance $R$ from the origin to the Gibbons--Hawking center,
\begin{equation}\label{eq:Rsol}
 R = a + \frac{16\,(x+1)^2}{(a^2- c^2)\,(q^1 q^2 q^3)^2+ 4\, ( q^1 q^2 + q^2 q^3 +q^3 q^1)\,(x+1)^2 }\,.
\end{equation}
The remaining vector fields in the metric all appear in the combinations 
\begin{eqnarray}  
\scal{ d\psi + w^0 }|_{r_1=0} &~=~&  d\psi + (N(1-\cos \theta_1)-1) d\varphi \,,  \CR
\Scal{ dy + \ax^3 ( d\psi + w^0 ) + w^3 }|_{r_1=0} &~=~& dy - \biggl(  \frac{ (a^2-c^2) q^1 q^2 q^3}{4 (x+1)} + \frac{x+1}{q^3}\biggr) ( d\psi - d\varphi) \,.\qquad\quad 
\end{eqnarray}
The second vector field is manifestly well-defined in the new coordinate 
\begin{equation} 
y_E = y - \biggl(  \frac{ (a^2-c^2) q^1 q^2 q^3}{2 (x+1)} +2 \frac{x+1}{q^3}\biggr)  \frac{ \psi - \varphi}{2}  \ , \label{yExtremal} 
\end{equation}
whereas the first is discontinuous. 

In these coordinates, considering the spacelike slice given by $dt=0$, the metric \eq{eq:6D-metr} becomes
\bea\label{GibbonsHawking}
ds^2 &=& \frac{h_3^E}{\sqrt{h_1^E h_2^E}} dy_E^{\; 2} \\
        && {} +N \sqrt{h_1^E h_2^E } \left[ r_1 \!\left(  d\!\!\;\left(\frac{\psi-\varphi}{N}\right) + (1-\cos \theta_1)d\varphi \right)^2 + \frac{dr_1^2}{r_1} + r_1 ( d\theta_1^2 + \sin^2\theta_1d\varphi^2) \right]  . \quad 
				\nonumber
\eea
%
Focusing temporarily on the second line in this metric, we recognize a Gibbons--Hawking self-dual metric.
If $(\psi-\varphi)$ had period $4\pi N$, this four-dimensional factor would simply be flat $\mathds{R}^4$, with $\frac{\psi-\varphi}{N}$ being the appropriately normalized Hopf fibre coordinate on $S^3$. 
Note that at infinity we have the identification $y \sim y+2\pi R_y$ at fixed $\psi$, $\varphi$. Thus in order for $(\psi-\varphi) \to (\psi-\varphi) + 4\pi N$ at fixed $y_E$ to be a closed orbit, in the change of coordinates \eqref{yExtremal} we require
\begin{equation} \label{eq:N3-def}
 -N\biggl(   \frac{ (a^2-c^2) q^1 q^2 q^3}{2 (x+1)} +2 \frac{x+1}{q^3} \biggr)  =N_3 R_y  
\end{equation}
for some integer $N_3$. 

Next, the periodicity of the $\psi$ and $\phi$ coordinates at infinity \eq{eq:identifications} means that the actual periodicity of $(\psi-\varphi)$ is $4\pi$ rather than $4\pi N$, and the resulting space can be thought of as arising via a $\mathds{Z}_{|N|}$ orbifold action\footnote{We often use the term `orbifold' in the physics sense, to denote a discrete quotient of a manifold. If the quotient results in no singularities, we denote it as a `smooth orbifold'.} on the smooth space described above. 
The $\mathds{Z}_{|N|}$ quotient acts on both $\psi$ and $y_E$ as
\begin{equation} \label{eq:orb-1}
\psi \rightarrow \psi + 4\pi \ , \qquad y_E \rightarrow y_E + \frac{2\pi N_3}{N} R_y \ , 
\end{equation}
and therefore the quotient is smooth provided $N_3$ and $N$ are relative primes. More generally the Euclidean base space has an orbifold singularity of degree ${\rm gcd}(N,N_3)$.

\subsection{Geometry at the bolt} 

The smoothness conditions at the bolt ($r=c$) are generalizations of those discussed in \cite{Bossard:2014ola} and in JMaRT \cite{Jejjala:2005yu}. At the bolt, the functions $W$, $H_I$ behave as
\begin{align}\label{eq:bolt-beh}
 W\big|_B = &\, \frac{\hat W_B(\theta)}{\sin^4\theta} \,, 
 \qquad
 H_I\big|_B =  \frac{\hat H_I^B(\theta)}{\sin^2\theta} \,,\qquad \mu\big|_B = \frac{\hat \mu_B(\theta)}{\sin^2\theta} \ , 
\end{align}
where we explicitly factor out the powers of $\sin\theta$ which diverge at the poles of the bolt. The $\hat W_B$, $\hat H_I^B $  and $\hat \mu_B$ are regular functions of $\theta$ on the bolt, whose explicit expressions are not very illuminating. Similarly, it is straightforward to compute that the gauge field components $\ax^I$ and $A^I_t$ are also regular functions at the bolt.

Turning to the vector fields, we consider the timelike fibration of the metric on the bolt, noting that \eqref{eq:Rsol} automatically ensures that the vector field $\omega$ is single-valued on the bolt. One may then compute the value of this field at $r=c$, to find 
\begin{align}\label{eq:omega-cons}
 \omega\big|_{B} = &\, \frac{a^2 -  c^2}{8\,a}\biggl( 2 \sum_J q^{J}-c\,  q^1q^2q^3 \biggr) 
   -4  \biggl( c \sum_I q^{I+1} q^{I+2} - 2 \biggr)  \frac{x^2 - 1}{8\,a\,q^1q^2q^3} 
 \\
 + &\, \left[ 16\,- (a + c)\, \left( (a^2 -  c^2)\,\frac{(q^1q^2q^3)^2}{(x+1)^2} + 4\,\sum_I q^{I+1} q^{I+2} \right)\right]\frac{(R-c)\,(x+1)\,N}{8\,a\,q^1q^2q^3\,(R-a)}\ ,\nonumber
\end{align}
which must vanish in order to avoid Dirac-Misner string singularities. Moreover, the function $\mu/W$ vanishes at the poles of the bolt due to \eqref{eq:bolt-beh}, so that the full vector field, $k$, vanishes on the symmetry axis. For $N=0$, the condition that \eqref{eq:omega-cons} be equal to zero reduces to the regularity constraint one gets in the JMaRT solution. In the following we will assume that \eqref{eq:omega-cons} vanishes, although it is preferable not to solve it explicitly yet. In practice we allow ourselves to define all quantities modulo terms proportional to  $\omega\big|_{B}$, that will eventually vanish once the constraint is solved explicitly. 


The vectors $w^0$ and $w^3$ are discontinuous on the symmetry axis at the poles of the bolt.  We consider therefore separately their value at the bolt (meaning in the limit $r\to c$) and their expression on the symmetry axis near the poles of the bolt. The values at the bolt are 
\begin{align}
  w^0\bigr|_B =&\, \frac{c}{a}\,x + \left(1+\frac{c}{a}\right)\,\frac{(R-c)}{(R-a)}\,N \,,
  \CR
  w^I\bigr|_B =&\,\frac{1}{a\, q^1 q^2 q^3}\Bigg[  \frac12\,(a^2 -  c^2) q^1 q^2 q^3 q^I - \biggl( c \,q^I \sum_{J\neq I} q^{J} - 2 \biggr) (x^2 - 1) \biggr .  
 \CR 
 &  \hspace{40mm} + \,  \biggl . \biggl( 4\,- (a + c)\, q^I \sum_{J\neq I} q^{J} \biggr) \, \frac{(R-c)\,(x+1)\,N}{(R-a)} \Bigg] \,,
\end{align}
and the values on the symmetry axis near the poles of the bolt are
\be\begin{split}
  w^0\bigr|_{\theta=\pi,\, r>c} &=1  \,, \\
  w^I\bigr|_{\theta=\pi,\, r>c}  &=0 \, , 
  \end{split}\hspace{10mm}\begin{split} 
w^0\bigr|_{\theta=0,\,c<r<R}  &= 2N-1\ , \\
w^I\bigr|_{\theta=0,\,c<r<R}  &= 2N \biggl(  \frac{ (a^2-c^2) q^1 q^2 q^3}{4 (x+1)} + \frac{x+1}{q^I}\biggr)   \ . 
\end{split}\ee
To interpret the discontinuities at $(r=c, \, \cos \theta =\pm1)$ we turn to the six-dimensional metric \eqref{eq:6D-metr}, in which both $w^0$ and $w^3$ appear explicitly. Considering a spacelike slice, $dt=0$, we find the following expression for the metric at the bolt
\begin{multline} \label{eq:6D-metraB}
ds^2\big|_{\text{bolt}} = \,\frac{\hat H^B_3}{\sqrt{\hat H^B_1 \hat H^B_2}}\,
\left( dy +  \ax^3 \,( d\psi + w^0|_B ) + w^3|_B  \right)^2 
\\
   \,        +  \frac{ \hat H_1^B  \hat H_2^B  \hat H_3^B  - \hat \mu^{\; 2}_B \sin^2 \theta}{\hat H_3^B \sqrt{\hat H_1^B \hat H^B_2 } \, \hat W_B} \, \sin^2 \theta \,\scal{  d\psi + w^0|_B }^2 \\
   \ + \sqrt{\hat H^B_1 \hat H^B_2 }\left[ a^2\left( \,\frac{dr^2}{r^2 - c^2  } + \, d\theta^2 \right) +  (r^2-c^2) \, d\varphi^2 \right] . \quad
\end{multline}
At the poles of the bolt (\ie $\cos \theta = \pm 1$), the second line in the above equation vanishes. Thus the only dependence of the metric on the vector field $d\psi + w^0$ is through the vector $ dy + \ax^3  (d\psi + w^0 ) + w^3$, which must be continuous at the poles.
Provided that \eqref{eq:omega-cons} vanishes, the (single-valued) limits of $\ax^3$ at the two poles satisfy
\begin{equation}
\ax^3 \big|_{(r=c,\cos \theta =\pm1)} = -\frac{ w^3\big|_{B} - w^3\big|_{\cos \theta=\pm 1,\,c<r<R} }{w^0\big|_{B} - w^0\big|_{\cos \theta=\pm 1,\,c<r<R}} \ ,  
\end{equation}
and therefore the vector field is indeed continuous at the two poles. 

The existence of a well-defined Killing vector $K_B$ defining the bolt requires the quantization of some of the parameters. To see this, it is convenient to introduce coordinates $\phi_-,r_B,\psi_-,\varphi_-$ in which the metric is manifestly well-defined on an open set excluding the North Pole $\theta=0$:
\begin{gather}
 y = R_y\, \phi_{-} \;,  \qquad r = c +\frac{1}{2c}\,r_B^{\; 2} \;, 
 \CR
\varphi = \varphi_{-} - (m-n)\, \phi_{-}  \;, \qquad 
\psi = \psi_{-}  - \varphi_{-} + (m+n)\, \phi_{-} \;,
\label{ChangeSouthPole} 
\end{gather}
where $m$, $n$ are the quantities appearing in the Killing vector \eqref{KillBolt} that degenerates at the bolt, which becomes $K_B = \partial_{\phi_-}$. For compatibility of the periodicities, $m$ and $n$ must be integers.
The quantities $a/c$ and $x$ are determined in terms of these integers as
\begin{equation} \label{eq:ax-mn}
\frac{a}{c} = m -n \ , \qquad  x +  \frac{(a + c)(R-c )}{c\,  (R-a)} N  = \frac{a}{c}\,w_\varphi^0\bigr|_B = m+n \;.
\end{equation}
In addition, the radius of the $y$ circle $R_y$ is fixed to
\begin{eqnarray} \label{eq:Ry-def}
R_y = \frac{a}{c}\, w_\varphi^3\bigr|_B &=& \frac{1}{c\,   q^1 q^2 q^3}\,
\biggl(  \tfrac12\,(a^2 -  c^2) q^1 q^2 (q^3)^2+\,\scal{  2-c \,q^3 (q^1+q^2)} (x^2 - 1) \biggr .  \hspace{20mm}
 \CR 
& & \hspace{22mm} + \,  \biggl . \biggl( 4\,- (a + c)\, q^3 (q^1 + q^2 ) \biggr) \, \frac{(R-c)\,(x+1)\,N}{(R-a)} \biggr)\,.
\end{eqnarray}
Given this value of $R_y$, we observe that the constraint \eqref{eq:N3-def} is a nontrivial constraint on the integers $N$ and $N_3$. 

The metric in the vicinity of the bolt then reduces to 
\begin{multline} 
ds^2\big|_{\text{bolt}} = \, 
\left(\frac{\hat H^B_3}{\sqrt{\hat H^B_1 \hat H^B_2}}\right)\, 
\left( \frac{R_y}{m-n} d\varphi_{-} + \ax^3\,\Scal{ d\psi_{-}   + \frac{2n}{m-n} d\varphi_{-} }   \right)^2 
\\
   \,     \qquad    +  \frac{ \hat H_1^B  \hat H_2^B  \hat H_3^B  - \hat \mu^{\; 2}_B \sin^2 \theta}{\hat W_B \hat H_3^B \sqrt{\hat H_1^B \hat H^B_2 }}  \sin^2 \theta \,\Scal{  d\psi_{-}   + \frac{2n}{m-n} d\varphi_{-} }^2 \\
   \ \qquad \qquad + \sqrt{\hat H^B_1 \hat H^B_2 }a^2\left(  \,\frac{dr^{\; 2}_B +r^{\; 2}_B  d\phi_{-}{}^2  }{c^2  } + \, d\theta^2  \right) \ .\label{eq:bolt-metr}
\end{multline}
The radial coordinate $r_B$ and the $2\pi$-periodic coordinate $\phi_{-}$ therefore parametrize $\mathbb{R}^2$ in radial coordinates, and the bolt metric is well-defined at a generic value of $\theta$.

\paragraph{Geometry at the poles of the bolt} \hspace{.2cm}\\
As mentioned earlier, the poles of the bolt are nuts. At the nuts, two $U(1)$ isometries collapse. In particular, the additional degenerate isometry at the South Pole $\theta=\pi$ follows from the fact that
\begin{equation}
 \ax^3\big|_{r=c,\theta=\pi} =  - \frac{R_y}{2n} \,
\end{equation}
with $R_y$ given by \eqref{eq:Ry-def}. This means that the additional degenerate isometry is along $\varphi_{-}$ and the leading dependence of the metric on $d\varphi_{-}$ as written in \eqref{eq:bolt-metr} vanishes (the subleading terms will appear in \eq{eq:SPmetric} below).
One then finds that in the neighbourhood of the South Pole, the geometry is the one of a regular Gibbons--Hawking nut times $S^1$, so the space is locally $S^1 \times \mathds{R}^4$.  To see this, we use the coordinates 
\be
r =\, \frac{1}{2} \Scal{ r_- + \sqrt{ r_-^{\; 2} - 4 c r_- \cos \theta_- + 4 c^2}} \,,
\quad~
\cos \theta = \, \frac{1}{2c} \Scal{ r_- -  \sqrt{ r_-^{\; 2} - 4 c r_- \cos \theta_- + 4 c^2}} \,,
\ee
and the constants
\begin{equation}
h_I^- = \frac{2\,(x+1)- (a+c)\, q^{I+1} q^{I+2}}{8\,c\,q^{I+1} q^{I+2}}
\left[ 2\,(x+2 N-1)-(a-c)\,q^{I+1} q^{I+2}+ \frac{4\,(a-c)\,N}{R-a} \right] ,  
\end{equation}
in terms of which the metric (with $dt=0$) in the neighborhood of $r_- = 0$ reduces to 
\bea\label{eq:SPmetric}
ds^2\big|_{r_-=0} &=&\, \frac{h^-_3}{\sqrt{h^-_1 h^-_2}}\,\left( \frac{R_y}{2n}d\psi_{-} \right) ^2  
\\ 
   & &\ \quad + \sqrt{h^-_1 h^-_2 }\left( \frac{1}{r_-} dr_-^{\; 2} + r_- \Scal{ d\theta_-^{\; 2}  + 2 (1-\cos \theta_- ) d\phi_{-}{}^2 + 2 (1+\cos \theta_-) d\varphi_{-}{}^2 }   \right) \ .\nonumber
\eea
This is manifestly a local product of $\mathbb{R}^4$ with an $S^1$ along $d\psi_{-}$, which remains finite in this limit.

In order to study the metric near the North Pole at $\theta=0$, one needs to change to a coordinate system that is regular there, unlike the coordinates in \eqref{ChangeSouthPole}. We therefore change to spatial coordinates $(\phi_{+},\psi_{+},r_B,\theta,\varphi_{+})$, where
\begin{equation} 
y = R_y ( \phi_{+} + N_3 \varphi_{+}  )  \ , \quad \psi = \psi_{+}   + (1-2N)  \varphi_{+} + (m+n) \phi_{+} \ , \quad \varphi = \varphi_{+} - (m-n) \phi_{+} \ , \label{ChangeNorthPole} 
\end{equation}
with $r_B$ defined in the same way as in \eqref{ChangeSouthPole} and $m$, $n$ still given by \eqref{eq:ax-mn}.
This gives
\begin{equation} 
\scal{  dy + \ax^3 \,( d\psi + w^0 ) + w^3 }\Big|_{r=c,\theta=0} ~=~ \frac{1+(m-n) N_3}{-2m+2(m-n) N} R_y d\psi_{+} \ , 
\end{equation}
and the degenerate isometry at the North Pole is associated to $\varphi_{+}$. 

In these coordinates, if $\phi_+$ is identified with period $2\pi$, then
$r_B,\, \phi_{+} $ define polar coordinates on $\mathds{R}^2$.
To examine the periodicities more closely, note that the change of coordinates  \eqref{ChangeNorthPole} is not unimodular, and its inverse is
\begin{gather} 
\phi_{+}  = \frac{ \frac{y}{R_y} - N_3 \varphi}{1+(m-n)\, N_3} \,, 
\qquad 
\varphi_{+} = \varphi + \frac{(m-n)\,(\frac{y}{R_y} - N_3 \varphi)}{1+(m-n)\, N_3}\ ,   
\CR
\psi_{+} = \psi + (2\,N-1) \varphi - 2\, \frac{(m-(m-n)\,N)\,(\frac{y}{R_y} - N_3 \varphi)}{1+(m-n)\, N_3} \,.
\end{gather}
We observe that $\psi_+$ has period $4\pi$, and that $\varphi_+$ has period $2\pi$. The periodicity of $y$ induces the identification 
\begin{gather} 
\!(\phi_+,\psi_+,\varphi_+) \sim (\phi_+,\psi_+,\varphi_+)+ 
2\pi\;\!\!\left(\frac{1}{1+(m-n) N_3}\,, - \frac{2(m-(m-n)N)}{1+(m-n) N_3}\,, \frac{(m-n)}{1+(m-n) N_3} \right)
\label{eq:orb-action}
\end{gather}
%
%
and we see that $\phi_+ \sim \phi_+ + 2\pi$ is contained in the full lattice of identifications.
Since $m-n$ and $1+(m-n)N_3$ are relatively prime for any integer $N_3$, the quotient is smooth at a generic point of the bolt, consistently with the fact that the metric was manifestly regular in the coordinates \eqref{ChangeSouthPole}. However, the coordinate $\varphi_{+} $ degenerates at the North Pole, and the orbifold action is only free at this point if $m-(m-n)N$ and $1+(m-n) N_3$ are relative primes.

To examine the geometry at the North Pole, consider the coordinate change 
\be
r =\, \frac{1}{2} \Scal{ r_+ + \sqrt{ r_+^{\; 2} + 4 c r_+ \cos \theta_+ + 4 c^2}} \,,
\quad~~
\cos \theta =\, \frac{1}{2c} \Scal{ \sqrt{ r_+^{\; 2} + 4 c r_+ \cos \theta_+ + 4 c^2}-r_+} \,.
\ee
%
We introduce the constants
\begin{equation} 
h_I^+ = \frac{2\,(x+1)+(a-c)\, q^{I+1} q^{I+2}}{8\,c \, q^{I+1} q^{I+2}}\,
\Scal{2\,(x+2 N-1)+(a+c)\,q^{I+1} q^{I+2} - \frac{4\,(a+c)\,N}{R-a}}\,,
\end{equation} 
in terms of which the metric (with $dt=0$) then reduces at $r_+\rightarrow 0$ to 
\bea
ds^2\big|_{r_+=0} &=&\, 
\frac{h^+_3}{\sqrt{h^+_1 h^+_2}} \left( \frac{(1+(m-n)N_3) R_y}{2m-2(m-n)N}d\psi_{+}\right)^2 
\\ 
   & &\ \quad + \sqrt{h^+_1 h^+_2 }\left( \frac{1}{r_+} dr_+^{\; 2} + r_+ \scal{ d\theta_+^{\; 2}  + 2 (1+\cos \theta_+ ) d\phi_{+}{}^2 + 2 (1-\cos \theta_+ ) d\varphi_{+}{}^2 }   \right) .\nonumber
\eea
Thus, similarly to the South Pole, the geometry at the North Pole is the product of a circle (parametrized by $\psi_{+}$) and a Gibbons--Hawking nut, with an orbifold action on $S^1 \times \mathds{R}^4$ of order $1+(m-n)N_3$ given by Eq.\;\eq{eq:orb-action}. 

\section{Topology and fluxes of the solutions}
\label{sec:top-flux}

In this section we discuss the topology and fluxes of our solutions. Before discussing the topology directly, we first examine the two-form potentials $B$ and $\tilde{B}$ near each of the nut centers $x_A$. 
We then discuss the two homology 3-cycles of the solution, and give explicit expressions for the fluxes associated to these two 3-cycles. 

\subsection{Two-form potentials at the nut centers}

We have seen in the last section that in the vicinity of each nut center $x_A$ (\ie the South Pole $x_-$ at $r_-=0$, the North Pole $x_+$ at $r_+=0$ and the extremal center $x_E$ at $r_1=0$), the 5-dimensional Riemannian base space is locally a smooth discrete quotient of $S^1\times \mathds{R}^4$, with the respective center as the origin of $\mathds{R}^4$ in adapted coordinates. 

In spherical coordinates, a regular 2-form on $\mathds{R}^4$ vanishes at the origin. In our solution, a regular 2-form must similarly reduce at each of the $x_A$ to its component along time and the $S^1$ that remains finite at that center. We have checked explicitly that the two-form potentials $B$ and $\tilde{B}$ evaluated at each center $x_A$ admit constant components in the base generated by $dt,\, dy,\, d\psi,\, d\varphi$ by the wedge product. It follows that one can define a gauge transformation such as to cancel the potential at the required point.  The relevant expressions at each center are rather long and not illuminating, so we refrain from displaying them. However, the difference of the values of the two-form at the centers carry information about the fluxes, as we now discuss in some detail.

Consider an open set, $U_A$, including the center $x_A$ and excluding the others, on which the regular 2-form potentials $B^\ord{A}$, $\tilde{B}^\ord{A}$, are defined as 
\begin{equation}\label{eq:reg-Bs}
B^\ord{A} \equiv B\big|_{U_A} - B\big|_{A} \, , \qquad 
\tilde{B}^\ord{A} \equiv \tilde{B}\big|_{U_A} - \tilde{B}\big|_{A} \,, 
\end{equation}
where $B|_A$ is $B$ evaluated at $x_A$, which defines a constant gauge transformation implementing \eqref{eq:reg-Bs}. On the intersection $U_{AB} \equiv U_{A}\cap U_B$, the two representatives are by construction patched modulo a gauge transformation
\begin{equation}\label{GaugeB}  
B^\ord{A}\big|_{U_{AB}}  - B^\ord{B}\big|_{U_{AB}}  = B\bigl|_B - B\bigl|_A \  , 
\end{equation}
and similarly for $\tilde B$. Explicitly, we find the following gauge transformations for the pullback of the 2-form, $B$, on a time-like slice $dt=0$:
\begin{align} \label{GaugeExplicit} 
 2\,\left( B \bigr|_{E} - B \bigr|_{N} \right) = &\,
 F_2\, \left( -N_3\,d\varphi\wedge d\psi -\frac{1}{R_y}\,(2\,N-1)\,d\varphi\wedge dy - \frac{1}{R_y}\,d\psi\wedge dy \right)\,,
\CR
 2\,\left( B \bigr|_{N} - B \bigr|_{S} \right) = &\,
 (Q_2+ N_3\,F_2)\,\left( d\varphi\wedge d\psi -\frac{m+n}{R_y}\,d\varphi\wedge dy - \frac{m-n}{R_y}\,d\psi\wedge dy \right)\,,
\end{align}
while the corresponding expressions for $\tilde B$ follow by exchanging the indices $1 \leftrightarrow 2$ in all expressions. Here $Q_a$ are the total electric charges  at asymptotic infinity \eqref{eq:Q-jmart} and $F_a$ define the fluxes 
\begin{align}\label{eq:F-def}
 \eta^{ab} F_b = &\,
 \frac{(a - c)\,(R - c)}{4 c\,(R - 2 a - c)}\,
 \frac{( (a + c)\, q^1 q^3 - 2\, (1 + x))\,((a + c)\, q^2 q^3 - 2\, (1 + x))}{(a - c)\, q^1 q^2 q^3 + 2\, (1 + x) q^a}
\CR
& \hspace{30mm}
\times \left( \frac{(a^2 - c^2)}{x + 1}\, q^1 q^2 q^3 + 2\, (a + c)\, (q^1 + q^2) + 4\, \frac{x - 1}{q^3} \right)\, . 
\end{align}
Note that the expressions of the gauge transformations become rather simple if we consider the coordinates at the extremal center \eqref{yExtremal} and  the South Pole \eqref{ChangeSouthPole}. Introducing $\psi_E$ as the coordinate in which the metric component $d \psi + w^0$ is well-defined on the axis between the North Pole and the extremal center, we obtain
\begin{align}
 2\,\left( B \bigr|_{E} - B \bigr|_{N} \right) = &\,
 F_2\,  \frac{1}{R_y} dy_E \wedge d\psi_E  \equiv \,
 F_2\,  \frac{1}{R_y}dy_E \wedge \scal{ d\psi + (2N-1) d\varphi} \,,
\CR
 2\,\left( B \bigr|_{N} - B \bigr|_{S} \right) = &\,
 (Q_2+ N_3\,F_2)\,d\varphi_- \wedge d\psi_-  \,.
\label{GaugeExplicit-2}
\end{align}

\subsection{Topology of the solutions}

We observe that only two linearly independent two-forms with integer coefficients appear in \eqref{GaugeExplicit} and \eq{GaugeExplicit-2}. This is a consequence of the presence of two inequivalent homology 3-cycles on any time-like slice of the solution. One can derive this fact from the Mayer--Vietoris sequence for the connected union of three spaces which are (smooth discrete quotients of) $S^1 \times \mathds{R}^4$. We define the five-dimensional Riemannian space  ${\cal M}_{n} $ through the recursive connected union (here $\cong$ denotes ``homeomorphic to''):
\begin{equation} \label{RecursiveSol} 
{\cal M}_{n+1} \cong {\cal M}_{n} \cup S^1 \times \mathds{R}^4 \ , \qquad  {\cal M}_{n} \cap S^1 \times \mathds{R}^4 \cong  S^1 \times S^1\times  \mathds{R}^2 \ , \qquad  {\cal M}_{0} \cong S^1 \times \mathds{R}^4  \ , 
\end{equation}
with the requirement that ${\cal M}_n$ is simply connected for $n\ge1$. Note indeed that the Riemannian base space of the solutions we describe in this paper are by construction simply connected because there is a basis in which each of the $U(1)$ isometries admits at least one fixed point.

The JMaRT solution is homeomorphic to ${\cal M}_1$, where the two  $S^1 \times \mathds{R}^4$ open sets are centered at the poles of the bolt, and a regular section of the bolt is indeed diffeomorphic to $S^1\times S^1\times \mathds{R}^2$, with the circles parametrized by $\psi_-,\, \varphi_-$. The bolt itself then defines a nontrivial 3-cycle, which can be viewed as a retraction of the $S^3$ present in the asymptotic $\mathbb{R}^{1,4}\times S^1$ region. For the union \eqref{RecursiveSol}, the Mayer--Vietoris sequence yields
\begin{equation} 
\dots \rightarrow H_{k}(S^1 \times S^1) \rightarrow H_k({\cal M}_n) \oplus H_k(S^1) \rightarrow H_k({\cal M}_{n+1}) \rightarrow H_{k-1}(S^1 \times S^1) \rightarrow \cdots 
\end{equation}
Setting $n=0$ we find for the JMaRT solution ${\cal M}_1$ the sequence
\begin{equation} 
0 \rightarrow H_3({\cal M}_1) \rightarrow \mathds{Z} \rightarrow 0 \rightarrow H_2({\cal M}_1) \rightarrow \mathds{Z}^2 \rightarrow \mathds{Z}\oplus\mathds{Z} \rightarrow 0 \ , 
\end{equation}
implying that $H_3({\cal M}_1)=\mathds{Z}$ and $H_2({\cal M}_1)$ is trivial, so one indeed finds that $\mathcal{M}_1 \cong \mathds{R}^2 \times S^3$, and we recover the nontrivial three-cycle of the JMaRT solution.

To analyze our solution, note that the effect of the pole in the function ${\cal H}_{\pA}$ in \eqref{eq:H-one} is to add an extra nut which is locally diffeomorphic to a smooth discrete quotient of $S^1 \times \mathds{R}^4$. It then follows that the solution displayed in this paper is homologically equivalent to ${\cal M}_2$ in \eqref{RecursiveSol}, obtained by one more recursion. The homology of our solution can thus be computed by the above Mayer--Vietoris sequence with $n=1$, which reads
\be 
0 \rightarrow \mathds{Z}\rightarrow H_3({\cal M}_2) \rightarrow \mathds{Z} \rightarrow 0 \rightarrow H_2({\cal M}_2) \rightarrow \mathds{Z}^2 \rightarrow \mathds{Z} \rightarrow 0 \,,
\ee
and implies that $H_3({\cal M}_2)=\mathds{Z}^2$, while $H_2({\cal M}_2)=\mathds{Z}$.
Thus our solution includes two inequivalent homology 3-cycles and one homology 2-cycle. Note that the details of the smooth discrete quotient of $S^1 \times \mathds{R}^4$ do not play any role in this construction. 

The Mayer--Vietoris sequence exhibits the isomorphism between the homology 3-cycles relating to centers $x_A$, and the $S^1\times S^1$ homology 2-cycle of the intersection $U_{AB}$, given by the restriction of the 3-cycle to the intersection. This isomorphism is dual to the isomorphism relating the cohomology representative 3-form $H$, and the gauge transformation \eqref{GaugeB} patching $B$ on $U_{AB}$. 

A similar analysis to the above can be performed in the five-dimensional bubbling black hole microstate solutions~\cite{Bena:2005va,Berglund:2005vb,Bena:2006kb,Bena:2010gg,Lunin:2012gp,Giusto:2012yz,Bena:2015bea}, where one can find that each additional Gibbons--Hawking center gives rise to an additional two-cycle.

\subsection{Fluxes on the 3-cycles}

One fundamental 3-cycle, $\Sigma_\infty$, is defined as the retraction of the asymptotic $S^3$ to the interior, as for the 3-cycle in the JMaRT solution. In our solution, it can be described as a surface with $dy=0$ and considering some path in $r,\theta$ coordinates from the South Pole to the Gibbons--Hawking center. Given that the intersections $U_{AB}$ of the three open sets are nontrivial, one would in principle need to consider a partition of unity in order to define the integrals for the fluxes. We avoid that by introducing a cellular complex $\{C_S\,,C_E\,,C_{SE}\}$ such that 
\begin{gather} 
C_S \subset U_S \ , \qquad C_E \subset U_E \ , \quad\quad  C_S \cap C_E  \cong \emptyset \ , \qquad C_{SE} \equiv \overline{C_S} \cap \overline{C_E} \ ,
\end{gather}
with
\begin{gather} 
U_S \cup U_E \cong C_S \cup C_E \cup C_{SE} \ , \qquad 
 C_{SE} \cong \partial C_S \cap  U_{SE} \cong \partial C_E \cap  U_{SE} \ ,
\end{gather}
so that one may replace $U_S$, $U_E$ by $\overline{C_S}$, $\overline{C_E}$ in all considerations, but with their intersection being retracted to the co-dimension one boundary $C_{SE}$. The integral of the three-form can then straightforwardly be computed as 
\bea 
\frac{1}{4\pi^2}  \int_{\Sigma_\infty} 
H &~=~	& 
\frac{1}{4\pi^2} \Bigg(\hspace{-2mm}
 \int\limits_{\quad\Sigma_\infty\cap C_S} \hspace{-2mm} d B^\ord{S} 
 ~+ \hspace{-3mm} \int\limits_{\quad\Sigma_\infty\cap C_E} \hspace{-2mm}
 d B^\ord{E} \Bigg)
 ~=~   \frac{1}{4\pi^2}\hspace{-4mm}
 \int\limits_{\quad\Sigma_\infty\cap C_{SE}} \hspace{-3mm}
   \left(B^\ord{S}- B^\ord{E} \right) \CR
&=&  \frac{1}{4\pi^2} \hspace{-4mm} \int\limits_{\quad\Sigma_\infty\cap C_{SE}}  
\!\! \left(B\big|_E- B\big|_S \right) ~=~ 
 \frac{1}{8\pi^2} Q_2 
 \int\limits d\varphi \wedge d\psi ~=~ Q_2 \ ,  
\eea
%
which gives the D5-brane charge. 
By construction, the integral of $\tilde{H}$ on the same cycle gives the D1-brane charge $Q_1$. 

The second fundamental 3-cycle $\Sigma_1$ can be defined in exactly the same way as the bubble linking the North Pole to the Gibbons--Hawking center on the axis while wrapping $\psi_E$ and $y_E$ and with $\varphi$ kept constant. One then computes the flux as
\bea 
\frac{1}{4\pi^2} \int\limits_{\Sigma_1} H &\;=\;& 
\frac{1}{4\pi^2} \Bigg( \hspace{-4mm} 
\int\limits_{~~~~\Sigma_1\cap C_N} \hspace{-2mm} d B^\ord{N}  
+ \hspace{-4mm} \int\limits_{~~~~\Sigma_1\cap C_E} 
\hspace{-2mm}	 d B^\ord{E} \Bigg)  
\;=\;  \frac{1}{4\pi^2} \hspace{-5mm} \int\limits_{~~~~\Sigma_1\cap C_{NE}}  
\!\!\! \left(B\big|_E-B\big|_N\right) ~=~ F_2 \ ,  \cr &&
\eea
with $F_2$ defined in \eqref{eq:F-def}, and similarly one finds that the flux of $\tilde H$ over $\Sigma_1$ is given by $F_1$. Dirac quantization therefore implies that the $F_a$ are quantized fluxes in appropriate units.

The integral of the three-form field strength over any cycle is a linear combination of $Q_2$ and $F_2$ with integer coefficients; this is ensured by \eqref{GaugeExplicit}. In particular, on the bolt itself (at $r=c$ and $dy=0$) one obtains
\be \label{eq:bolt-flux}
\frac{1}{4\pi^2} \int_{\Sigma_B} H = Q_2 + N_3 F_2 \ . 
\ee
A similar expression holds for the integral of $\tilde{H}$. 

The charges $Q_1$ and $Q_2$ are quantized in string theory as follows. Taking a $T^4$ compactification of type IIB for concreteness, we consider $n_1$ D1-branes wrapped on the $y$-circle $S^1_y$, and $n_5$ D5-branes wrapped on $T^4 \times S^1_y$. Then denoting the volume of the $T^4$ at infinity by $(2\pi)^4 V$ and the string coupling by $g_s$, the supergravity charges take the standard form (see \eg\cite{Peet:2000hn})
\be
Q_1 ~=~ \frac{g_s n_1 \alpha'^3}{V} \,, \qquad Q_2 ~=~ g_s n_5 \alpha' \ .
\ee
The flux $F_1$ is quantized in the same way as the D1 charge, and similarly the flux $F_2$ is quantized in the same way as the D5 charge.

We thus observe the familiar story that these solitonic solutions are supported by fluxes, as discussed in~\cite{Gibbons:2013tqa} and also by \cite{deLange:2015gca,Haas:2014spa,Haas:2015xmc}.
It would be interesting to verify explicitly that the Komar-type integral defining the mass of our solution can be decomposed using the intersection form of the Euclidean base space, as discussed in~\cite{deLange:2015gca}.
In addition, one could examine the analogous formulae for the angular momenta as Komar-type integrals. We anticipate that this could be used to 
show that flux quantization implies angular momentum quantization in the appropriate units.


\section{Explicit examples of smooth solutions}
\label{sec:num-example}

We now present explicit examples of smooth solutions of the type described in the previous two sections. While most of the regularity and smoothness constraints have been imposed analytically above, there remain three regularity conditions to be solved.

The first of the remaining regularity conditions is the condition that $\omega$ vanish at the bolt, $\omega\bigr|_{B}=0$, where $\omega\bigr|_{B}$ is given in Eq.\;\eq{eq:omega-cons}.
Our second constraint comes from the regularity condition at the bolt, which relates $x$ to $m+n$, given in Eq.\;\eq{eq:ax-mn}.
The third constraint is that \eq{eq:N3-def} must be solved by some integer $N_3$.

A priori, one would wish to fix the integer $N_3$ first and solve for the remaining parameters,
however the constraints cannot be solved analytically in terms of $N_3$ and the other integer parameters $m$, $n$, $N$. Therefore we take a different approach: we first solve for $N_3$ in terms of the other parameters, and then verify that the parameter space allows examples where $N_3$ is an integer.

To simplify the three constraints, we use the second condition \eq{eq:ax-mn} to eliminate $N$ from $\omega\bigr|_{B}$ in favour of $m$, $n$ and $x$.
Conveniently, this happens to also eliminate $R$ from $\omega\bigr|_B$. The first regularity condition then becomes
\begin{align}\label{eq:omega-cons-2}
& \hspace{-3mm} (a^2 -  c^2) q^1q^2q^3 \biggl( 2 \sum_J q^{J}-c\,  q^1q^2q^3 \biggr) -4  \biggl( c \sum_I q^{I+1} q^{I+2} - 2 \biggr)  (x^2 - 1) 
 \CR 
 & {}\, =\, \left[ 16\,- (a + c)\, \left( (a^2 -  c^2)\,\frac{(q^1q^2q^3)^2}{(x+1)^2} + 4\,\sum_I q^{I+1} q^{I+2} \right)\right]\frac{(x+1)\big(x-(m+n)\big)}{(m-n+1)}\,,
\end{align}
where we retain $a$ in some places for ease of notation, but it should be understood that $a$ takes the value $(m-n)c$ from \eq{eq:ax-mn}.

We next eliminate $R$ from the second condition \eq{eq:ax-mn}, using \eq{eq:Rsol}. The second condition then becomes
\bea\label{eq:x-con-2}
&&{} 
x-(m+n)+(m-n+1)N ~=~ \cr 
&&{} 
\qquad\quad c\,N\,\left((m-n)^2-1\right)
\frac{(a^2- c^2)\,(q^1 q^2 q^3)^2+ 4\, ( q^1 q^2 + q^2 q^3 +q^3 q^1)\,(x+1)^2 }{16\,(x+1)^2} \,. \qquad
\eea
We thus have two polynomial constraints on the parameter space, \eqref{eq:omega-cons-2} and \eqref{eq:x-con-2}, which we choose to solve for the variables $q^3$ and $x$. 
To solve these two polynomials simultaneously for $q^3$, we take the resultant with respect to $q^3$, which (after removing overall factors) gives a quartic in $x$ depending on $c$, $m$, $n$, $q^1$, $q^2$, $N$.
The full quartic would take more than a page to write out, and is not particularly illuminating, so we do not reproduce it here. 

In the limit $N \to 0$, this quartic has a double root at $x=m+n$ and another double root. We focus on the two roots which tend to the JMaRT value $m+n$ in the $N \to 0$ limit.
We thus obtain $x$ in terms of $c$, $m$, $n$, $q^1$, $q^2$, $N$; since this is a solution to a complicated quartic, the answer obtained is algebraically very complex.

Next, the constraint \eqref{eq:omega-cons-2} is quadratic in $q^3$, enabling us to solve for $q^3$ as a function of 
$c$, $m$, $n$, $q^1$, $q^2$, $N$, $x$. We again select the root which joins smoothly to the JMaRT solution.

Given the algebraic complexity involved, we investigate the regularity of the solution by scanning the parameter space numerically, as follows. The dimensionful parameter $c$ merely sets the scale of the system, so we work in units of $c$. 
Using the first two conditions, specifying values for $m$, $n$, $q^1$, $q^2$, $N$ determines in turn $x$, then $q^3$. Then $R$ is determined from \eq{eq:Rsol} and $N_3$ is given by the third regularity condition~\eq{eq:N3-def}.
Given such a set of parameters, we examine the remaining regularity conditions that away from the special points $H_I$ are positive and finite, and that Eqs. \eq{eq:cond1} and \eq{eq:cond2} are satisfied.

Next, we investigate whether the parameter space allows $N_3$ to be an integer. To do so, we first find a region of parameter space that satisfies all other regularity checks, and we then tune one of the parameters to make $N_3$ come within some desired precision of an integer value.

As discussed around \eq{eq:orb-1} and \eq{eq:orb-action}, the geometry will be free of orbifold singularities if ${\rm gcd}(N,N_3)=1$ and ${\rm gcd}(m-(m-n)N, 1+(m-n) N_3)=1$. We now describe such an example, which is completely smooth.

From our numerical investigations, we did not find any smooth solutions with $|N| \le 3$, however for $N=-4$ we found a region of parameter space that allows regular solutions. In this region it appears that $N_3$ can be tuned to be as close as desired to the quantized value $N_3 =3$; we find a region in which $N_3$ is within $10^{-8}$ of this value, both above and below the quantized value.

A representative example of a solution is given by:
\bea
\quad  m=3 \,, \quad n=1 \,,\quad  N = -4 \,, \quad q^2 = 0.5 \,c^{-\frac12} \,, \quad q^1 = 0.672558 \,c^{-\frac12}
\eea
where $N_3 = 3 \pm 10^{-8}$ has been approximately quantized by tuning $q^1$.

This solution is well-behaved everywhere: both the orbifold actions \eq{eq:orb-1} and \eq{eq:orb-action} are smooth quotients. In addition to being smooth at the special points, away from these points it satisfies the regularity conditions  discussed in Section \ref{sec:reg-general}. 

Let us now describe some of the properties of the solution.
The regime of small parameters $q^I$ corresponds to the regime of large supergravity charges $Q_I$. Ultimately $Q_I$ should be thought of as macroscopic, but in our example we have kept the numbers relatively modest in units of $c$ for convenience. Rounding smaller quantities to three significant figures and larger quantities to integers, the values of some quantities of interest in this solution are:
\setlength\arraycolsep{6.5pt}
\be
\begin{array}{lllll}
	x = 62.3\,, &  q^3 = 12.8\,c^{-\frac12} \,, & R = 2.25\,c\,, & R_y = 13.3\,c^{\frac12} \,, \\
  Q_1 = 2392\,c\,, & Q_2 = 1767\,c\,, & Q_3 = 46156\,c\,, &M_{\mbox{\tiny ADM}} = 50408\, c \ ,   \\
  F_1 = -482\, c\, , & F_2 = -364\, c\, ,& J_\psi = 452034\,c^{\frac32} \,,  & J_\varphi = 53503\,c^{\frac32} \, .
\end{array}
\ee

Comparing to the regularity bound on angular momenta for a black hole carrying the charges $Q_I$ and the mass $M_{\mbox{\tiny ADM}}$,
we find that the angular momentum $J_\varphi$ is below the regularity bound, while $J_\psi$ is slightly over-rotating. To understand this, note that the behavior of the solution at infinity is determined by the charges $Q_I$ and the constants $E_I$  \eqref{eq:E-jmart}, which by \eqref{ConE} are themselves determined by the charges and the ADM mass. A formal black hole solution with the same charges, angular momenta and mass would have an entropy $S_{\mbox{\tiny BH}} = S_{\mbox{\tiny L}} + S_{\mbox{\tiny R}}$, where \cite{Cvetic:1996kv,Cvetic:2011hp}
\bea \label{Jbounds} 
\biggl( \frac{S_{\mbox{\tiny L}}}{2\pi}\biggr)^2 \hspace{-4mm} &=&   \hspace{-2mm}  \frac{1}{8} \Bigl({ \scriptstyle \sqrt{ (E_1 +Q_1)(E_2+Q_2)(E_3+Q_3)} \;+\;   \sum\limits_I \sqrt{ (E_I +Q_I)  (E_{I+1} -Q_{I+1}) (E_{I+2} -Q_{I+2})}} \Bigr)^2 - J_\psi^{\; 2} \ , \CR 
 \biggl( \frac{S_{\mbox{\tiny R}}}{2\pi}\biggr)^2  \hspace{-4mm} &=&\hspace{-2mm}   \frac{1}{8} \Bigl({ \scriptstyle \sqrt{ (E_1 -Q_1)(E_2-Q_2)(E_3-Q_3)} \;+\;   \sum\limits_I \sqrt{ (E_I -Q_I)  (E_{I+1} +Q_{I+1}) (E_{I+2} +Q_{I+2})}} \Bigr)^2 - J_\varphi^{\; 2} \ . \hspace{7mm}  \eea
In the BPS limit, $E_I \to Q_I$, and the above formula reduces to the familiar BMPV cosmic censorship bound~\cite{Breckenridge:1996is}.
Using the expressions derived in Section \ref{sec:reg-general}, one obtains 
\bea \label{JboundsSol} \biggl( \frac{S_{\mbox{\tiny L}}}{2\pi}\biggr)^2 \hspace{-4mm} &=&   \hspace{-2mm}  - 4\biggl( (a^2-c^2)(q^1 + q^2 + q^3) + 4 \frac{x^2-1}{q^1 q^2 q^3} \biggr)^2  \ , \CR 
 \biggl( \frac{S_{\mbox{\tiny R}}}{2\pi}\biggr)^2  \hspace{-4mm} &=&\hspace{-2mm}   -\biggl( c \biggl( (a^2 - c^2) q^1 q^2 q^3 + 4\sum_I \frac{x^2-1}{q^I}\biggr)  + \frac{ 16 N (x+1) (R^2-c^2)}{q^1 q^2 q^3 (R-a)^2}\biggr)^2  \CR
 && \hspace{5mm} - \frac{32 (a-c) (x+1) N (R^2-c^2)}{q^1 q^2 q^3 (R-a)^2} \biggl(  (a^2 - c^2) q^1 q^2 q^3 + 4 \sum_I \frac{x^2-1}{q^I}\biggr)  \ . \hspace{8mm}  \eea
Therefore the solutions described in this paper necessarily have $J_\psi$ exceeding the cosmic censorship bound, whereas $J_\varphi$ can possibly preserve the bound for a negative $N$. In our example $J_\varphi$ is below the regularity bound and $S_{\mbox{\tiny R}} = 499 712 \, c^\frac{3}{2}$, whereas $J_\psi$ exceeds the regularity bound by a rather small amount, 
\bea
\frac{J_\varphi^{\; 2} }{ J_\varphi^{\; 2} +(\frac{S_{\mbox{\tiny R}}}{2\pi})^2 } ~\approx~ 0.31 \,,
\qquad\qquad 
\frac{J_\psi^{\; 2} }{ J_\psi^{\; 2} +(\frac{S_{\mbox{\tiny L}}}{2\pi})^2 } ~\approx~ 1 + \frac{1}{62^2} \;.
\eea
Note moreover that Eq.\;\eqref{JboundsSol} is only valid within the specific solution \eqref{eq:LK-gen} discussed in this paper, and is not a general property of solutions to the partially solvable system defined on the Maxwell--Einstein instanton background  \eqref{eq:Kerr-mod}. Within our understanding, there is no reason to believe that the over-rotation is a general property of solutions to the system.

The over-rotation is to be expected; it is a feature which is also present in the JMaRT solutions, and one may expect that adding a single center in a simple way would not change this fact. Note nonetheless that the JMaRT solutions have the two angular momenta exceeding the regularity bound, so that the addition of an extra Gibbons--Hawking center is an improvement in this respect. In addition, we observe that the ADM mass is above but quite close to the BPS bound $\sum_I Q_I = 50315 \, c$. 

The ergoregion of the six-dimensional solution is larger than the scales of the charges $Q_I$, and extends until around $r = 11554\,c$. By contrast, the would-be ergoregion of the five-dimensional solution obtained upon reduction along the $y$ fiber is much smaller, extending to around $r = 169\, c$; this is consistent with the fact that one regularity bound is satisfied, and the other violated only weakly. 
The difference can be traced to the fact that the momentum charge in the $y$ direction $Q_3$ is significantly larger than the D1 and D5 charges $Q_1$ and $Q_2$.
In the JMaRT solutions, in the near-BPS limit the ergoregion is deep inside an AdS$_3 \times S^3$ throat; interestingly, as a result of the large ergoregion in the six-dimensional solution, there is no such throat in our solution.


\section{Discussion}
\label{sec:disc}

In this paper we have constructed solutions to six-dimensional $\cN=(1,0)$  supergravity coupled
to a tensor multiplet that are the first non-extremal smooth horizonless solutions containing
both a bolt and an additional Gibbons--Hawking center. This center lies at a fixed distance
from the bolt, giving rise to two inequivalent 3-cycles supported by three-form flux.
These solutions are generalizations of the JMaRT solutions \cite{Jejjala:2005yu}, and reduce to them upon removing the additional Gibbons--Hawking center.

Our solutions have an asymptotic structure similar to that of non-extremal black holes in five dimensions, albeit with one of the two angular momenta exceeding the regularity bound for black holes. The fact that $J_{\varphi}$ is under-rotating and that $J_{\psi}$ is over-rotating only by a very small amount is a significant improvement compared to the JMaRT solutions, for which both angular momenta are over-rotating. In the context of the fuzzball proposal, our solutions should be viewed as describing atypical semi-classical microstates of non-extremal D1-D5-P black holes.

At the level of the system of equations, there does not appear to be anything to indicate that this atypicality should be a general feature of all solutions to this system. 
Rather, it is a common feature of explicitly-constructible microstate geometries that their fluxes tend to produce angular momenta larger than those of black holes, so to have $J_{\varphi}$ under-rotating in this solution is a noteworthy feature.
For supersymmetric multi-center solutions involving $N_c$ Gibbons--Hawking centers, the ratio between the square of $J_\psi$ and the product of the charges has been estimated to be equal to one plus corrections of order $1/N_c^2$~\cite{Bena:2006is}. 
In our solution the equivalent correction is approximately $1/62^2$. One can think of the bolt as corresponding to two Gibbons--Hawking centers, so in some sense our solution can be thought of as having three centers, and thus the amount of over-rotation appears remarkably small.

In the future one would of course like to make the further improvement of obtaining solutions that have both angular momenta within the black hole regime.
The only known way to do this is to consider specific multi-center solutions in which one can tune the fluxes in order to make the distance between the centers arbitrarily small~\cite{Bena:2006kb,Bena:2007qc}. One refers to these solutions as scaling solutions~\cite{Bates:2003vx}. 
Such microstate geometries play an important role in the fuzzball proposal, as they naturally admit an arbitrary long throat, and have been argued~\cite{Bena:2006kb} to be dual to typical states of the D1-D5 orbifold CFT~\cite{Vafa:1995zh}.

It is an exciting possibility that there may also exist a scaling regime for solutions far from the BPS limit, and indeed far from extremality. In this case there would not be an AdS throat, and the relevant physical parameter should be the redshift between the locus of the centers and the asymptotic region, which could possibly be tuned to become arbitrary large as the centers approach each other in the supergravity approximation. 

To obtain scaling non-extremal solutions, the first necessary ingredient is of course to add more centers, and in this paper we have given a proof of principle that this can be done.
Our solutions are not however in the scaling regime, and have a large ergoregion which is not contained inside an AdS$_3 \times S^3$ throat. 
Earlier experience with BPS solutions suggests that it is difficult to construct axisymmetric scaling solutions with less than four centers \cite{Bena:2006kb,Vasilakis:2011ki}. 
Since our solution can be thought of as having three centers, if we had found scaling behavior it would have been surprising.

In principle, it is straightforward to use our methods to construct solutions with an arbitrary number of centers, despite the complexity of the relevant equations. 
To obtain our solutions, we worked in a restricted ansatz which explicitly disallows interaction between the extremal centers.
We expect that, upon turning on such interactions, one can obtain solutions with enough Gibbons--Hawking centers to allow for a scaling behavior in the sense described above. In the near-BPS limit, we expect that it should be possible to obtain a large AdS$_3 \times S^3$ throat encompassing all centers and any ergoregion.

It is an important problem to understand the stability of our solutions and their possible microscopic interpretation. It is well-known that the JMaRT solutions are unstable to decay via ergoregion emission~\cite{Cardoso:2005gj}. Since in the near-BPS limit these solutions have large AdS$_3 \times S^3$ regions, this instability can be studied holographically. In this limit the ergoregion is deep inside the throat, and in the dual CFT the ergoregion emission is naturally interpreted as the Hawking radiation emitted by the dual CFT states~\cite{Chowdhury:2007jx,Avery:2009tu,Avery:2009xr}.
Until recently, the dual states had been known for only a subset of parameters of the full JMaRT solutions, however recently the dual CFT states of the most general JMaRT solutions have been identified~\cite{Chakrabarty:2015foa}, and the emission spectrum and rate have been found to match between gravity and CFT for all parameters.
While our present solutions do not appear to have standard AdS$_3 \times S^3$ throats, they do have ergoregions, and thus one may also expect them to decay via ergoregion emission. It would be interesting to investigate the corresponding decay rate and emission spectrum.

There has been recent work which constructs JMaRT solutions using inverse scattering techniques~\cite{Katsimpouri:2014ara}. These methods also offer the prospect of building multi-center generalizations of JMaRT, and may provide a complementary line of enquiry to that described here.

Looking further to the future, it would be interesting to investigate the relationship between our results and an interesting recent proposal involving long-string degrees of freedom at the inner horizon of non-extremal black holes~\cite{Martinec:2014gka,Martinec:2015pfa}. More generally, it would be interesting to gain further insight into how large a subset of the degrees of freedom of non-extremal black holes can be described within supergravity.

Our construction of non-extremal multi-bubble microstate geometries represents a long-sought-after technical advance, which we anticipate will enable the construction of many more non-extremal solitonic supergravity solutions involving topological cycles supported by flux, and thereby provide a deeper understanding of the quantum physics of non-extremal black holes.

\vspace{0.5cm}

\section*{Acknowledgements}

We would like to thank Bert Vercnocke and especially Nick Warner for valuable discussions. SK is grateful to CPHT, Ecole Polytechnique, for hospitality in the final stages of this work. The work of IB and DT was supported by John Templeton Foundation Grant 48222 and by a grant from the Foundational Questions Institute (FQXi) Fund, a donor advised fund of the Silicon Valley Community Foundation on the basis of proposal FQXi-RFP3-1321 (this grant was administered by Theiss Research). The work of SK was supported in part by INFN and by the ERC Grant 307286 (XD-STRING). The work of DT was also supported by a CEA Enhanced Eurotalents Fellowship.

\vspace{0.5cm}


\begin{appendix}

\section{Relation to the 5D and 4D ansatze}
\label{app:reduction}

In this appendix we give a few comments on the reduction of the system solving six dimensional $\cN=(1,0)$ supergravity described in Section \ref{sec:sys-ans} to five and four dimensional supergravity with eight supercharges.  

Upon reduction on the circle parametrized by $y$ in \eqref{eq:6D-metr}, one obtains $\cN=1$ supergravity in five dimensions, coupled to two vector multiplets. There are three gauge fields in the theory, with one belonging in the supergravity multiplet, appearing completely symmetrically in the action. One is the Kaluza--Klein gauge field, $A^3$ in \eqref{eq:6D-metr}, while the other two arise by reduction of the two dual two-forms, $B$ and $\tilde B$ as
\begin{align}
 B = &\, ( dy + A^3 ) \wedge A^1 + B_2 \ ,
 \CR
 \tilde{B} =  &\, ( dy + A^3 )  \wedge A^2 + B_1\,.
\end{align}
Here, the two-form fields $B_a$ are dual to the field strengths of the $A^a$ in five dimensions, \ie   
\begin{align}\label{eq:B2-dual}
\Bigl( \frac{H_1^{\; 2}}{H_2 H_3}\Bigr)^{\frac{2}{3}} \star_5 F^1  =&\, dB_1 + A^2 \wedge F^3 \ ,
 \CR
\Bigl( \frac{H_2^{\; 2}}{H_1 H_3}\Bigr)^{\frac{2}{3}} \star_5 F^2  =&\, dB_2 + A^1 \wedge F^3 \ ,
\end{align}
which follows from the six dimensional equation of motion \eqref{eq:6D-eom}. With these definitions, one finds that the gauge fields are given by
\begin{align}
A^I = &\, A^I_t\, (dt + \omega) + \ax^I\,(d\psi + w^0) + w^I\,,
\end{align}
for $I=\{1,2,3\}$ and the components $A^I_t$, $\ax^I$ and $w^I$ are given by \eqref{eq:5dzeta}--\eqref{eq:5dax} and \eqref{wIEq-1}.

The further reduction along the isometry described by the angle $\psi$ in \eqref{eq:6D-metr} leads to four dimensional $\cN=2$ supergravity coupled to three vector multiplets. Now, there are four gauge fields, $A^\Lambda$, for $\Lambda=\{0,I\}=\{0,1,2,3\}$, with one belonging to the supergravity multiplet. The reduction of the particular system of equations studied in this paper from five to four dimensions was briefly discussed in the Appendix of \cite{Bossard:2014ola}, so we focus on the direct translation of the
four dimensional quantities into the six dimensional quantities of Section \ref{sec:sys-ans}.

The metric in four dimensions takes the form
\begin{gather}\label{eq:metric4-gen}
  ds_4^2= - e^{2U} (dt + \omega)^2 + e^{-2U} \gamma_{ij} dx^i dx^j  \,,
\CR
 e^{-4U} = W^{-1} \left(H_1 H_2 H_3 - \mu^2 \right)\,,
\end{gather}
while the gauge fields and their electromagnetic duals are given by
\begin{equation}
 A^\Lambda = \zeta^\Lambda (dt + \omega) + dw^\Lambda\,,\hspace{10mm}  A_\Lambda = \zeta_\Lambda (dt + \omega) + dv_{\Lambda}\,.
\end{equation}
Here, the $dw^\Lambda$ are given by \eqref{wIEq-1} and \eqref{wIEq-2}, while the two of the dual $dv_{\Lambda}$ are given in \eqref{eq:alm-NE-mag}, with the remaining ones
\begin{align} \label{eq:alm-NE-mag-2}
\star d v_0 =&\,  2\,V\, (\Pp  d \cE_- - \cE_- d \Pp ) + 2 \, \cE_- \Pp  \star d \sigma\,, 
\CR
\star d v_3 =&\, V\, (\cE_- d \cE_+ - \cE_+ d \cE_-) - \cE_+ \cE_- \star d \sigma\ ,
\end{align}
being automatically conserved due to the Ernst equations. The four-dimensional complex scalars are given by
\begin{equation}\label{eq:4dscal}
 z^I =  \alpha^I + \mathrm{i}\, \frac{e^{-2U}}{H_I} \,,
\end{equation}
where the axions $\alpha^I$ are given in \eqref{eq:5dax}.

In view of \eqref{eq:2-form-exp}, it is straightforward to lift any solution of four dimensional supergravity to six dimensions, once the fields $db_a$, $\beta_a$, $A^a_t$ are given in terms of four dimensional quantities. The scalars $A^a_t$ are given by
\begin{align}\label{eq:4dzeta}
 \zeta^0 = - e^{4U} \mu \,,
 \qquad
 A^I_t = &\, \zeta^I + \ax^I \,\zeta^0\,,
\end{align}
where we also give the timelike component of the Kaluza--Klein gauge field $A^3$ and the function $\mu$ in \eqref{eq:6D-metr} in terms of the $\zeta$'s. The remaining six dimensional quantities are given by
\begin{align}
db_1 + \frac{H_1^{\; 2}\,W}{H_1 H_2 H_3-\mu^2 } *_3 d \ax^1 = &\, \zeta^0 dv_1 + \zeta_1 dw^0 - \zeta^3 dw^2 - \zeta^2 dw^3 
+ \left( \zeta^0 \zeta_1 - \zeta^3 \zeta^2 \right) d\omega \,,
\CR
db_2 + \frac{H_2^{\; 2}\,W}{H_1 H_2 H_3-\mu^2 } *_3 d \ax^2 = &\, \zeta^0 dv_2 + \zeta_2 dw^0 - \zeta^3 dw^1 - \zeta^1 dw^3 
+ \left( \zeta^0 \zeta_2 - \zeta^3 \zeta^1 \right) d\omega\,,
\end{align}
and
\begin{align}\label{eq:beta6d}
\beta_1 = -( \zeta_1 + a^2 \zeta^3 )\,, 
\qquad
\beta_2 = -( \zeta_2 + a^1 \zeta^3 )\,.
\end{align}
Conversely, one may invert \eqref{eq:4dzeta}--\eqref{eq:beta6d} to obtain the four dimensional scalars $\zeta^\Lambda$, $\zeta_1$, $\zeta_2$, without the need to pass through the five dimensional theory. For completeness, we give the final two components of the $\zeta_\Lambda$:
\begin{align}
\zeta_3 = &\, \frac14\, V\, \cE_- \left(\cE_+^2 (K_3 + V\, \Pm L^1 L^2 + V\, K_a L^a) + 2\, \Pp\, K_1 K_2 \right. 
\CR
         &\,\left.  \hspace{5cm}   -\cE_+ (\Pp L^3 + K_a L^a + 2\, V \,\Pp K_1 K_2)  \right) \,, 
\CR
\zeta_0 = &\, \frac12\, V\, \cE_- \left( \Pp\, (\Pp L^3 - \cE_+ K_3) - \Pp (1 + V\,\cE_+) \,K_a L^a 
              +  2\, V\, \Pp^2 K_1 K_2  \right. 
\CR
         &\,\left.  \hspace{7cm}  + \cE_+\, (2 - V\,\Pm\, \Pp)\, L^1 L^2\right)\,. 
\end{align}

\section{Expressions for vector fields}\label{app:vecfields}

The explicit expressions for the vector fields are given in terms of the conserved vector currents used in the construction of the solution in \cite{Bossard:2014ola}. The relevant basis is given by
\begin{align} \label{eq:currs}
 \mathcal{J}_0 = &\, \frac{dr}{r^2 - c^2 + a^2 \sin^2\theta}  
   + \frac{2\, a^2 \cos{\theta}\, (\cos{\theta}\, dr - r\, d\cos{\theta} ) }{(r^2 - c^2 + a^2 \sin^2\theta)^2}\,,
\CR
 \mathcal{J}_1 = &\, \frac{d\cos{\theta}}{r^2 - c^2 + a^2 \sin^2\theta}  
   + \frac{2\,r\, (\cos{\theta}\, dr - r\, d\cos{\theta} ) }{(r^2 - c^2 + a^2 \sin^2\theta)^2}\,,
\CR
 \mathcal{J}_2 = &\, \frac{ \cos{\theta}\, dr - r\, d\cos{\theta} }{(r^2 - c^2 + a^2 \sin^2\theta)^2}\,,
\CR
 \mathcal{J}_3= &\, d \mathcal{H} 
 + \frac{c^2-a^2}{m_-}\,\frac{ d \mathcal{H}\,( r-a \cos{\theta}+m_- )-\mathcal{H}\, d ( r-a \,\cos{\theta}) }{ r^2-c^2 + a^2 \sin^2\theta }
\CR
&\, 
+2\,a\,\frac{c^2-a^2}{m_-}\,\mathcal{H}\,(r -a \cos{\theta}+ m_- )\frac{ \cos{\theta}\, d r- r\,d \cos{\theta} }{\left(r^2-c^2 + a^2 \sin^2\theta\right)^2}\,,
\CR
 \mathcal{J}_4=&\, 
 \frac{ d \mathcal{H}\,\scal{  r+a \cos{\theta}+\frac{c^2-a^2}{m_-} }+\mathcal{H}\,d ( r+a \,\cos{\theta}) }{ r^2-c^2 + a^2 \sin^2\theta }
\CR
&\, 
+ 2\,a\, \mathcal{H} \,\Scal{ r +a \cos{\theta} +\tfrac{c^2-a^2}{m_-} }\frac{ \cos{\theta}\, d r - r\,d \cos{\theta} }{\left(r^2-c^2 + a^2 \sin^2\theta\right)^2}\,,
\end{align}
which define the associated vector fields through $\mathcal{J}_\gimel = \star d W_\gimel$, as 
\begin{align} \label{eq:currs-2}
W_0 = &\,  -\frac{(r^2 - c^2)\cos\theta }{r^2 - c^2 + a^2 \sin^2 \theta}  d\varphi  \,,
\CR
W_1 = &\,\frac{r \sin^2 \theta }{r^2 - c^2 + a^2 \sin^2 \theta} d\varphi  \,,
\CR
W_2 = &\,\frac{1}{2} \frac{ \sin^2 \theta }{r^2 - c^2 + a^2 \sin^2 \theta} d\varphi  \,,
\CR
W_3= &\, \mathcal{H} \frac{c^2 - a^2}{m_-} \biggl( \cos \theta + 
      a \sin^2\theta \frac{ r - a \cos\theta + m_-}{r^2 - c^2 + a^2 \sin^2 \theta} \biggr)d\varphi    \CR
      &\qquad + \sum_\pA \cH_\pA  \biggl(  a + 
      r \cos\theta  - \frac{(R_\pA^{\; 2} - c^2)( r + a \cos \theta + \tfrac{c^2-a^2}{m_-})}{ ( R_\pA-a) r + ( a R_\pA - c^2 ) \cos \theta } \biggr) d\varphi \ ,
 \CR
W_4=&\, \mathcal{H} \biggl( \frac{ a \sin^2 \theta  \scal{ r + a \cos \theta + \tfrac{c^2-a^2}{m_-}}}{r^2 - c^2 + a^2 \sin^2 \theta} - \cos\theta \biggr) d\varphi  \CR
& \qquad   + \sum_\pA  \cH_\pA  \frac{ r^2 - c^2 \cos^2 \theta + \frac{c^2-a^2}{m_-} ( r - R_\pA \cos \theta ) }{(R_\pA-a)r + ( a R_\pA - c^2 ) \cos \theta } d\varphi 
\, . 
\end{align}
We therefore give the relevant coefficients for each of the vector fields, employing a five-com\-ponent vector notation, so that the vectors given below should be contracted with the vector \\ $(W_0,\,W_1,\,W_2,\,W_3,\,W_4\,)$:
\begin{align}
 w^0= &\, \frac{1}{4} \,
      \begin{pmatrix}
       e_- (p^3+2 q^3) - m_- (l_a q^a+l^3)+\frac{1}{2} q^1 q^2 (m_+-m_-)\\
       \\
       -a\, e_- (p^3+2 q^3) + a\,m_- (l_a q^a+l^3)+\frac{1}{2} a\, q^1 q^2 (m_-+m_+) \\
       \\
       -2 \,a\,(a^2-c^2 )\, (l_a q^a+l^3)-a\,q^1 q^2 (a^2-c^2) \\
       -2\,a\,e_- m_+ p^3 + a\,m_- m_+ q^1 q^2 \\
       \\
       -\frac{1}{2} q^1 q^2\\
       \\
       \frac{1}{2}\,m_{-} q^1 q^2 - e_- p^3
      \end{pmatrix}\,,
\end{align}
\begin{align}
 w^1= &\, \frac{1}{4} \,
      \begin{pmatrix}
       -2\, e_- p^1-q^1 (m_-+m_+) \\
       \\
       -2\,a \, e_- p^1 - a \,q^1 (m_--m_+) \\
       \\
        2\,a \,q^1 \left(a^2-c^2+m_- m_+\right) + 4\,a \,e_- m_+ p^1 \\
        \\
        q^1\\
        \\
       2\, e_- p^1 + m_- q^1
      \end{pmatrix}\,,
\end{align}
\begin{align}
 w^3= &\, \frac{1}{4} \,
      \begin{pmatrix}
 \frac{1}{e_-}\,(a^2-c^2-m_-^2 )\, (l_a q^a+l^3) \\ -2\, e_- p^1 p^2+ (m_+-m_-) (p_a q^a + p^3)-2\,m_- q^3 \\
 \\
 \frac{a}{e_-}\,(a^2-c^2-m_-^2 )\, (l_a q^a+l^3) \\ + 2\,a\,e_- p^1 p^2 + a\,(m_-+m_+) (p_a q^a + p^3) + 2\,a\,m_- q^3  \\
 \\
 4\,a\,\frac{m_-}{e_-} \,(a^2-c^2)\, (l_a q^a+l^3)-2\,a\, (a^2-c^2-m_- m_+ )\, (p_a q^a + p^3)
  \\
  +4\,a\,q^3 \,(a^2-c^2) - 4\,a\,e_- m_+ p^1 p^2 \\
 \\
   -p_a q^a - p^3\\
   \\
   2\, e_- p^1 p^2 + m_-(p_a q^a + p^3)
      \end{pmatrix}\,,
\end{align}
\begin{equation}
 \omega= \frac{1}{8} \,
      \begin{pmatrix}
 \frac{1}{e_-}\,(a^2-c^2-m_-^2 )\, (l_a q^a+l^3) + \frac{1}{e_-} \,(a^2-c^2 + m_- m_+ )\,q^1 q^2 \\ 
 + 2\, e_- p^1 p^2+ (m_+ + m_-) (p_a q^a + p^3) + 2\,m_- q^3 \\
 \\
 -\frac{a}{e_-}\,(a^2-c^2+m_-^2 )\, (l_a q^a+l^3) -\frac{a}{e_-} \,(a^2-c^2 + m_- m_+ )\,q^1 q^2 \\ 
 + 2\,a\,e_- p^1 p^2 - a\,(m_+ -m_-) (p_a q^a + p^3) + 2\,a\,m_- q^3  \\
 \\
 2\,a\, (a^2-c^2 + m_- m_+ )\, (-p_a q^a + p^3) -\frac{2\,a\,m_-}{e_-} \,(a^2-c^2 + m_- m_+ )\,q^1 q^2
  \\
  +4\,a\,q^3 \,(a^2-c^2) + 4\,a\,e_- m_+ p^1 p^2 \\
 \\
   -p_a q^a - p^3 - \frac{m_-}{e_-}\,q^1 q^2 \\
   \\
   -2\, e_- p^1 p^2 + m_-(-p_a q^a + p^3) - \frac{m_-^2}{e_-}\,q^1 q^2
      \end{pmatrix}\,.
\end{equation}

\end{appendix}

\newpage

\bibliography{PaperG} \bibliographystyle{JHEP}
 
\end{document}